\PassOptionsToPackage{table}{xcolor} 
\documentclass[letterpaper,twocolumn,10pt]{article}
\usepackage{usenix2024_SOUPS}

\usepackage[T1]{fontenc} %

\usepackage{pdflscape}
\usepackage{array}
\usepackage{makecell}
\usepackage{multirow}
\usepackage{booktabs}
\usepackage[para, flushleft]{threeparttable}
\usepackage{tabularx}
\usepackage{colortbl}
\newcolumntype{L}[1]{>{\raggedright\arraybackslash}p{#1}} %
\newcolumntype{C}[1]{>{\centering\arraybackslash}p{#1}} %
\newcolumntype{R}[1]{>{\raggedleft\arraybackslash}p{#1}} %

\usepackage{float} %
\usepackage{subcaption}

\usepackage[inline]{enumitem}

\usepackage{amsmath} 
\usepackage{amssymb}

\usepackage{xargs} %
\usepackage{xstring} %

\usepackage[%
    autopunct=true,
]{csquotes} %

\DeclareQuoteStyle[american]{english} %
        {\itshape\textquotedblleft}
        [\textquotedblleft]
        {\textquotedblright}
        [0.05em]
        {\textquoteleft}
        {\textquoteright}

\usepackage{tikz}
\usetikzlibrary{shapes,arrows, arrows.meta,matrix, positioning, fit, calc, decorations.pathreplacing}

\newbool{displaycomments}
\setbool{displaycomments}{true}

\ifbool{displaycomments}{%
    \definecolor{springgreen}{rgb}{0.0, 1.0, 0.5}
    \definecolor{dandelion}{HTML}{f0e130}
    \newcommand\COMMENTX[2]{{\bf({#1}: {#2})}}
    \newcommand\sascha[1]{\COMMENTX{sascha}{{{\color{orange}#1}}}}
    \newcommand\yas[1]{\COMMENTX{yas}{{{\color{pink}#1}}}}
    \newcommand\jan[1]{\COMMENTX{jan}{{{\color{red}#1}}}}
    \newcommand\byron[1]{\COMMENTX{byron}{{{\color{purple}#1}}}}
    \newcommand\florian[1]{\COMMENTX{florian}{{{\color{green}#1}}}}
    \newcommand\marco[1]{\COMMENTX{marco}{{{\color{blue}#1}}}}
    \newcommand\angela[1]{\COMMENTX{angela}{{{\color{green}#1}}}}
    \newcommand\supervisor[1]{\COMMENTX{for supervisor}{{{\color{purple}#1}}}}
    \newcommand\jacques[1]{\COMMENTX{jacques}{{{\color{brown}#1}}}}
    \newcommand\todo[1]{\COMMENTX{todo}{{{\color{red}#1}}}}
    \newcommand\draft[1]{{{\color{gray}#1}}}
    \newcommand\red[1]{{{\color{red}#1}}}
}{%
    \newcommand\sascha[1]{}
    \newcommand\lea[1]{}
    \newcommand\yas[1]{}
    \newcommand\jan[1]{}
    \newcommand\byron[1]{}
    \newcommand\florian[1]{}
    \newcommand\marco[1]{}
    \newcommand\angela[1]{}
    \newcommand\supervisor[1]{}
    \newcommand\jacques[1]{}
    \newcommand\todo[1]{}
    \newcommand\draft[1]{}
    \newcommand\red[1]{}
}

\newcommand{\boldparagraph}[1]{\paragraph{#1}}
\makeatletter
\renewcommand\boldparagraph{\@startsection{paragraph}{4}{0\parindent}%
    {0.6ex plus 0.6ex minus 0.2ex}%
    {0ex}%
    {\normalfont\normalsize\bfseries\maybe@addperiod}%
}
\newcommand{\maybe@addperiod}[1]{%
    \let\@period\@empty%
    \def\@IEEEsectpunct{}%
    #1\@addpunct{.}\enspace%
}
\makeatother

\usepackage[usetwoside=true]{mdframed}
\usepackage{tcolorbox}
\tcbuselibrary{breakable}
\tcbuselibrary{skins}
\usepackage{enumitem}
\definecolor{darkgray}{gray}{0.3}
\definecolor{middlegray}{gray}{0.7}
\tcbset{}
\newtcolorbox{summaryBox}[2][]
{
    enhanced,
    breakable,
    frame hidden,
    borderline west = {3pt}{0pt}{middlegray},
    colback         = white,
    size            = fbox,
    left            = 0.3em,
    enlarge top by  = 0.3em,%
    coltitle        = black,
    title           = {\color{darkgray} \textbf{#2.} },
    attach title to upper,
    fontupper=\small, 
    #1,
}

\newlength\bubblesize
\newcommand{\yes}{\tikz[baseline=0.1ex] \fill[black]
(\bubblesize,\bubblesize) circle (\bubblesize);}

\newcommand{\no}{\tikz[baseline=0.1ex] \draw[black, line width=0.2ex] (\bubblesize,\bubblesize) circle (\bubblesize-.5\pgflinewidth);}

\newcommand{\definevar}[2]{%
  \expandafter\newcommand\csname var#1var\endcsname{#2}%
}
\newcommand{\var}[1]{\ifcsname var#1var\endcsname%
        \csname var#1var\endcsname%
    \else\PackageWarning{Var}{`#1' does not exist.}%
        \red{TODO}%
    \fi%
}

\usepackage[
    backend=biber,
    isbn=false,
    doi=false,
    url=false,
    eprint=true,
    defernumbers=true,
    useprefix=true, %
    mincitenames=1,
    maxcitenames=2,
    maxbibnames=9,%
    mincrossrefs=99,%
    uniquelist=false,
    style=numeric
    ]{biblatex}

\DeclareCiteCommand{\citeauthorfull}
    {\boolfalse{citetracker}%
    \boolfalse{pagetracker}%
    \DeclareNameAlias{labelname}{given-family}%
    \usebibmacro{prenote}}
    {\ifciteindex
        {\indexnames{labelname}}
        {}%
        \printnames{labelname}}
        {\multicitedelim}
{\usebibmacro{postnote}}

\usepackage{url}

\usepackage[acronym]{glossaries}

\newglossaryentry{fp}{%
    name={citing publication},
    description={} %
}

\newglossaryentry{fps}{%
    name={citing publications},
    description={} %
}
\newglossaryentry{FP}{%
    name={Citing Publication},
    description={} %
}
\newglossaryentry{FPS}{%
    name={Citing Publications},
    description={} %
}

\newglossaryentry{pets}{%
    name={PETS/PoPETs},
    description={} %
}
\newglossaryentry{cscw}{%
    name={ACM CSCW},
    description={} %
}
\newglossaryentry{usenix}{%
    name={USENIX Security},
    description={} %
}
\newglossaryentry{hci-cpt}{%
    name={HCI for Cybersecurity, Privacy and Trust},
    description={} %
}

\newacronym{usp}{USP}{usable security and privacy}
\newacronym{nlp}{NLP}{natural language processing}
\newacronym{soups}{SOUPS}{Symposium on Usable Privacy and Security}
\newacronym{hci}{HCI}{human--computer interaction}
\newacronym{sp}{S\&P}{security and privacy}

\definevar{papers.soups}{27}
\definevar{papers.fws}{26}
\definevar{fws.total}{129}
\definevar{fws.median}{4}
\definevar{fws.std}{2.67}
\definevar{fws.words.avg}{39.25}
\definevar{fws.words.median}{35}
\definevar{fws.words.std}{20.47}
\definevar{fws.implemented}{8}
\definevar{fws.implemented.text}{eight}
\definevar{fws.implemented.author}{3}
\definevar{fws.implemented.author.text}{three}
\definevar{fws.implemented.others}{6}
\definevar{fws.implemented.others.text}{six}
\definevar{fps.implemented.author}{1}
\definevar{fps.implemented.author.text}{one}
\definevar{fps.implemented.conferences}{4}
\definevar{fps.implemented.conferences.text}{four}

\definevar{papers.fw.implemented}{5}
\definevar{papers.fw.implemented.text}{five}%
\definevar{papers.fps.implementing-fws}{7}
\definevar{papers.fps.implementing-fws.text}{seven}%
\definevar{papers.amount.soups-with-one-future-paper}{3}
\definevar{papers.amount.soups-with-one-future-paper.text}{three}
\definevar{papers.amount.soups-with-two-future-paper}{2}
\definevar{papers.amount.soups-with-two-future-paper.text}{two}

\definevar{papers.count.future.work.statements}{7}%
\definevar{papers.count.future.work.statements.text}{seven}

\definevar{replication.amount}{6}
\definevar{replication.amount.text}{six}%

\definevar{citations.all}{1,484}
\definevar{papers.citing.all}{1,148} %
\definevar{citations.following}{1,207}
\definevar{papers.fps}{978} %

\definevar{citations.downloaded}{1,387}

\usepackage{orcidlink}

\begin{document}

\date{}

\title{How the Future Works at SOUPS: Analyzing Future Work Statements and Their Impact on Usable Security and Privacy Research}

\author{
    {\rm Jacques Suray}\,\orcidlink{0009-0007-8595-3706}\textsuperscript{$\ast$}
        \and
    {\rm Jan H.\ Klemmer}\,\orcidlink{0000-0002-6994-7206}\textsuperscript{$\mathcal{C}$}
        \and
    {\rm Juliane Schmüser}\,\orcidlink{0000-0001-7830-6403}\textsuperscript{$\mathcal{C}$}
        \and
    {\rm Sascha Fahl}\,\orcidlink{0000-0002-5644-3316}\textsuperscript{$\mathcal{C}$}\\[-12pt]
        \and
    \textsuperscript{$\ast$}Leibniz University Hannover, Germany, \hypersetup{hidelinks}\texttt{\href{mailto:suray@stud.uni-hannover.de}{suray@stud.uni-hannover.de}}\\
    \textsuperscript{$\mathcal{C}$}CISPA Helmholtz Center for Information Security, Germany, \\
    \hypersetup{hidelinks}\texttt{\{\href{mailto:jan.klemmer@cispa.de}{jan.klemmer},\href{mailto:juliane.schmueser@cispa.de}{juliane.schmueser},\href{mailto:sascha.fahl@cispa.de}{sascha.fahl}\}@cispa.de}
}

\def\plainauthor{Jacques Suray, Jan H. Klemmer, Juliane Schmüser, Sascha Fahl}

\maketitle

\begin{abstract}

Extending knowledge by identifying and investigating valuable research questions and problems is a core function of research.
Research publications often suggest avenues for future work to extend and build upon their results.
Considering these suggestions can contribute to developing research ideas that build upon previous work and produce results that tie into existing knowledge.
Usable security and privacy researchers commonly add future work statements to their publications. 
However, our community lacks an in-depth understanding of their prevalence, quality, and impact on future research.

Our work aims to address this gap in the research literature.
We reviewed all \var{papers.soups}~papers from the 2019 SOUPS proceedings and analyzed their future work statements.
Additionally, we analyzed \var{papers.fps}~publications that cite any paper from SOUPS 2019 proceedings to assess their future work statements' impact. 
We find that most papers from the SOUPS 2019 proceedings include future work statements.
However, they are often unspecific or ambiguous, and not always easy to find.
Therefore, the \gls{fps} often matched the future work statements' content thematically, but rarely explicitly acknowledged them, indicating a limited impact. 
We conclude with recommendations for the usable security and privacy community to improve the utility of future work statements by making them more tangible and actionable, and avenues for future work.

\end{abstract}

\section{Introduction}
\label{sec:intro}

Peer-reviewed scientific publications\footnote{\textbf{Disclaimer:} We use the terms \textit{scientific publications}, \textit{research articles}, and \textit{papers} interchangeably in the rest of this work.} are critical to communicating research findings to other researchers and the general public.
When extending the overall body of knowledge with novel insights, researchers frequently build upon and work in the context of prior research, e.g., developing new ideas, confirming or falsifying prior results, or comparing their work with related previous research.
Moreover, researchers read many papers or search journals and conference proceedings for new scientific insights, and also to get inspiration for future research. 
Scientific publications often suggest avenues for future work to extend and build upon their results.
We refer to those suggestions as \emph{future work statements} in the rest of this paper.
In recent years, there have been some efforts to analyze future work statements in the \gls{nlp} community~\cite{journal/informetrics/ZhangXHLQW23, conf/asist/ZhuWS19}, and information system design researchers have examined future work statements in their field to propose a research debt life cycle~\cite{journal/PAJAIS/BarataC23}.
\Gls{usp} researchers also regularly add such statements to their publications.

Despite this common practice, the \gls{usp} community lacks an in-depth understanding of the reporting of future work statements, and how future research addresses and implements these statements.
It is unclear how often such statements are included, what they contain, and how specific they are.
Guidelines for paper content and writing, such as the \gls{soups} call for papers~\cite{soups-cfp} or the \citeyear{how-to-soups} article on how to write a \gls{soups} publication~\cite{how-to-soups}, do not address future work statements specifically.
Moreover, the question arises of how future work statements in \gls{soups} research articles affect future research and thereby the overall development of the field.
To address the above gap in the \gls{usp} research literature, we investigate the following research questions in the course of this paper:\\

\begin{description}\itemsep0em
    \item[RQ1] \textit{How do \gls{soups} research articles include future work statements?}
        When reading research articles, researchers commonly encounter statements about ideas for future research.
        In this work, we aim to better understand the content and specificity of future work statements in \gls{soups} research articles.
    \item[RQ2] \textit{To what extent do researchers address future work statements from SOUPS research articles?}
        A key aspect of future work statements is the stimulation of new research ideas and projects.
        We aim to gain in-depth insights into the contribution of future work statements to the future of \gls{usp} research.
    
\end{description}

To answer these research questions, we first analyze all \var{papers.soups}~publications from the \gls{soups} 2019 proceedings.
We identify and extract all future work statements and explore their content and quality.
Second, for the papers that contained future work statements, we analyzed \var{papers.fps}~ research articles that cite those to check whether and how the initial future work statements are addressed by \gls{fps}.

In our work, we make the following contributions:

\begin{description}\itemsep0em
    \item[Analysis of Future Work Statements:] 
        We analyzed \var{fws.total} future work statements in the \gls{soups} 2019 proceedings for their content, location, and specificity. We find that suggestions for new research topics or goals are most common, followed by calls to investigate potential influence factors for observed phenomenons. Many future work statements are broad, ambiguous, or have a very large scope.
        
    \item[Analysis of \gls{FPS}:]
        For the \var{papers.fws}~publications that contained future work statements, we investigate \var{papers.fps}~\gls{fps}, to analyze how those implement future work statements. 
        We find that only \var{papers.fps.implementing-fws.text}~publications implemented a total of \var{fws.implemented.text} future work statements.
        Our findings suggest that the impact of future work statements on inspiring future \gls{usp} research is limited.

    \item[Recommendations:] 
        Based on our findings, we make suggestions for the \gls{usp} community to improve future work statements and their utility (\autoref{sec:discussion-recommendations}).  
        We argue that future work statements do not always add value, but if present should be specific, easy to find, and acknowledged by research they influenced.
        
    \item[Dataset:]
        We publish the dataset of annotated future work statements and the \gls{fps} as an artifact online (see \nameref{sec:availability} section).
\end{description}

\section{Related Work}\label{sec:related-work}

We discuss related work in two key areas: 
(i)~research regarding future work statements,
and 
(ii)~meta-research in the usable security and privacy community.

\boldparagraph{Research on Future Work}
Research on context-based citation analysis and use of information retrieval and \gls{nlp} techniques in bibliometrics investigates future work statements.
In 2014,~\citeauthor{conf/asist/DingZCSWZ14} presented a study on content-based citations, which addresses citation interpretation on a syntactic and semantic level. 
They contrasted content analysis as the traditional, manual approach, and \gls{nlp} as an emerging automatic or semi-automatic approach~\cite{conf/asist/DingZCSWZ14}. 
\citeauthor{journal/NLE/JhaJQR17} presented their \gls{nlp}-driven analysis approach in 2016, aiming to aggregate the scientific community's attitude towards a given publication as a measure of impact~\cite{journal/NLE/JhaJQR17}.
An essay by \citeauthor{conf/sigir/Teufel17} illustrates challenges and opportunities in evaluating the impact of future work statements using \gls{nlp}~\cite{conf/sigir/Teufel17}.
In 2019, \citeauthor{conf/issi/LiY19} proposed a keyword-based approach to identify future work statements, to improve the integration of future work statements into quantitative studies of science~\cite{conf/issi/LiY19}. 
\citeauthor{conf/jcdl/HaoLQWZ20} followed up on this with the creation of a manually curated data set of annotated future work statements from \gls{nlp} conference papers~\cite{conf/jcdl/HaoLQWZ20}.
Recently, researchers used \gls{nlp} models to identify key insights~\cite{arxiv/MarshalovaBB23}, research contributions~\cite{journal/scientometrics/ChaoCZL23}, and future work statements~\cite{journals/corr/HuW15a} in scientific publications.
Automatically identified future work statements were further utilized to identify research trends~\cite{conf/asist/QianLHWZ21}, and generate novel research ideas~\cite{journal/DADK/RuoxuanLY21}.
\citeauthor{journal/informetrics/ZhangXHLQW23} calculated the correlation of future work statements with abstracts of subsequent publications in the \gls{nlp} domain, 
finding that similarity increased with later years of publication of the future work statements. 
They additionally observed that similarity typically decreased the more years had passed after the publication of the future work statement, but sometimes also increased again at later points, suggesting a resurgence of a research topic~\cite{journal/informetrics/ZhangXHLQW23}.
A content analysis of the future work statements in \gls{nlp} research found four types; supportive, methodological, identifying potential influence factors, and presenting future targets~\cite{conf/asist/ZhuWS19}.
In a literature review of research into venue recommenders, \citeauthor{journal/DTA/DehdariradGJ20} found future work statements commonly suggested the use of more datasets or new algorithms~\cite{journal/DTA/DehdariradGJ20}.
\citeauthor{journal/PAJAIS/BarataC23} conducted a manual analysis of future work statements in information system design, finding that many focus on the paper's limitations. 
They introduce a research debt life cycle, showcasing points in the research process where future work statements can and do have an impact~\cite{journal/PAJAIS/BarataC23}.

\gls{nlp}-based automatic approaches are still an area of active research and existing models do not necessarily generalize.
We chose to manually analyze future work statements for this first analysis of future work statements and their impact in \gls{usp}.

\boldparagraph{Meta-Research in Usable Security and Privacy and Adjacent Fields}
Meta-research to investigate research and publication practices is an upcoming topic in the \gls{usp} community.
In 2021, \citeauthor{journal/tochi/DistlerFHKLLCK21} investigated the representation of risk in security studies with human participants. 
They also report on topics studied and methods used in \gls{usp} publications from 2018--2022, and provide guidelines to improve the clarity and comprehensiveness of reporting~\cite{journal/tochi/DistlerFHKLLCK21}.
In 2023, \citeauthor{conf/usenix/HasegawaIA24} reviewed 715~\gls{usp} papers to analyze the demographics of populations that are studied in \gls{usp} research. 
They found that participant demographics are often not reported, and if reported skew heavily towards WEIRD (Western, Educated, Industrialized, Rich, and Democratic) countries~\cite{conf/usenix/HasegawaIA24}.

We add our insights on the prevalence and impact of future work statements to this growing body of knowledge about meta-research in \gls{usp} and adjacent fields.

\section{Methodology}
\label{sec:method}

\begin{figure}[htb!]
    \centering
    \footnotesize
    \begin{tikzpicture}[%
        auto,
        block_main/.style ={rectangle, draw=black, fill=none, text width=0.9\columnwidth, text ragged, minimum height=3em, inner sep=3pt, text centered},
        block_double/.style ={rectangle, draw=black, fill=none, text width=0.41\columnwidth, text ragged, minimum height=11em, inner sep=3pt},
        block_double_dashed/.style ={rectangle, dashed, draw=black, fill=none, text width=0.41\columnwidth, text ragged, minimum height=8em, inner sep=3pt},
        block_noborder/.style ={rectangle, draw=none, fill=none, minimum height=17.25em, inner sep=0pt, text width=0.41\columnwidth, text centered},
        line/.style ={draw, thick, -latex', shorten >=0pt}]
    ]
	\begin{scope}[node distance = 0.3cm]
        \node [block_main] (proceedings) {
            \textbf{\gls{soups} 2019 Proceedings}\\
            Collected all \var{papers.soups}~publications.
        };
        \node [block_double_dashed, below=of proceedings.south west, anchor=north west] (identify-fws) {
            \textbf{1a. Identify Future Work Statements}\\
            Manually reviewed all publications to identify future work statements.
        };
        \node [block_double_dashed, below=of proceedings.south east, anchor=north east] (collect-fps) {
            \textbf{1b. Collect Citations}\\
            Collected all \var{citations.all} citations of the \gls{soups} 2019 papers via Google Scholar.
        };
        \node [block_noborder, below=of identify-fws] (placeholder) {
        };
        \node [block_double, below=of placeholder] (fws) {
            \textbf{2a. Future Work Statements}\\
            \var{papers.fws} \gls{soups} 2019 papers contained \var{fws.total}~future work statements.
        };
        \node [block_double_dashed, below=of collect-fps] (filter-fps) {
            \textbf{2b. Filter for \gls{FPS}}\\
            Applied exclusion criteria: 
            \begin{itemize}
                \setlength\itemsep{.1em}
                \item[--] no/dead links: 29
                \item[--] paywalls: 45
                \item[--] non-English: 76
                \item[--] non-publication formats: 18
                \item[--] broken PDFs: 22
                \item[--] cite paper without statements: 50
                \item[--] duplicates: 266
            \end{itemize}
        };
        \node [block_double_dashed, below=of fws] (analyze-fws) {
            \textbf{3a. Future Work Statement Analysis}\\
            Analyzed future work statements for content, location, and specificity.
        };
        \node [block_double, below=of filter-fps] (fps) {
            \textbf{3b. \gls{FPS}}\\
            \var{papers.fps}~papers cited \var{papers.fws} \gls{soups} 2019 containing future work statements papers \var{citations.following} times.\footnote{Some \gls{fps} cited multiple 2019 SOUPS papers.}
        };
        \node [block_double_dashed, below=of fps] (analyze-fps) {
            \textbf{4b. Analyze \gls{FPS}}\\
            Analyzed if and how a \gls{fp} implements a future work statement.
        };
 
    \end{scope}
    \begin{scope}[every path/.style=line]
        \path ($(proceedings.south west)!0.47!(proceedings.south)$) -- (identify-fws.north);
        \path ($(proceedings.south)!0.53!(proceedings.south east)$) -- (collect-fps);
        \path (identify-fws) -- (fws);
        \path (collect-fps) -- (filter-fps);
        \path (fws) -- (analyze-fws);
        \path (filter-fps) -- (fps);
        \path (fps) -- (analyze-fps);
        \path (fws.north east) -- (filter-fps.south west);
        \path (fws.south east) -- (analyze-fps.north west);
    \end{scope}
    \end{tikzpicture}
\caption{Summary of our two-phase systematic literature analysis of 27 papers in the SOUPS 2019 proceedings and their \var{papers.fps} citing research articles.}
\label{fig:method-overview}
\end{figure}
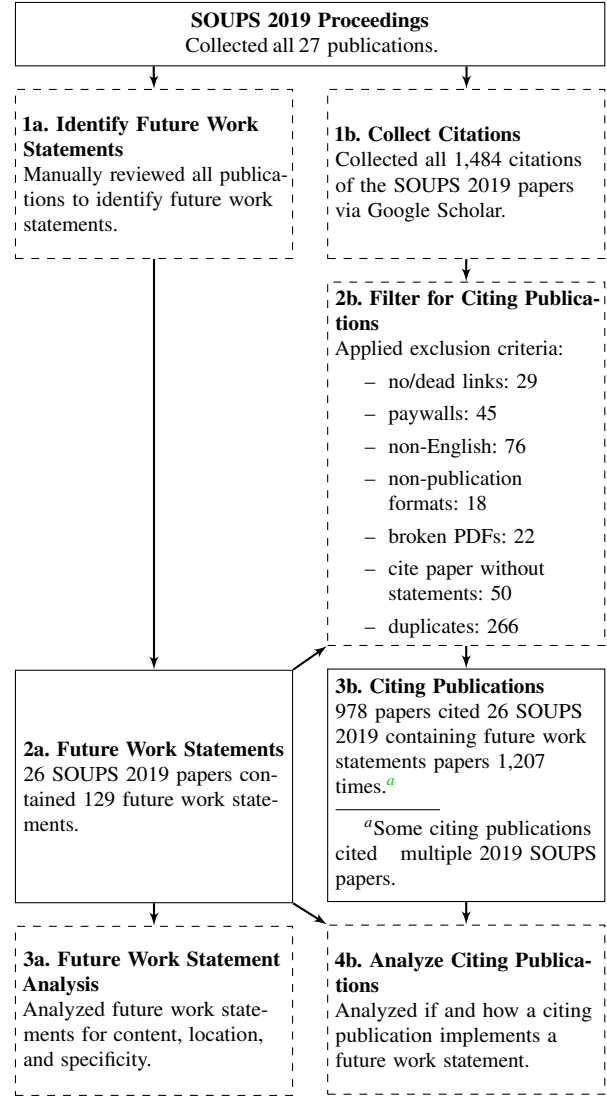

Below, we describe our approach to the systematic literature review of future work statements and their impact on \gls{usp} research.
Our analysis consists of two parts, as depicted in the methods overview in \autoref{fig:method-overview}.
(i)~First, we investigate all \var{papers.soups}~publications from the \gls{soups} 2019 proceedings to identify future work statements and analyze them in depth.
(ii)~Second, based on these publications, we analyzed whether and how \var{papers.fps}~publications that cite one of the \gls{soups} 2019 publications implement or address the originally proposed future work.
We decided to focus on \gls{soups} publications, as \gls{soups} is the premier venue with a focus on publishing \gls{usp} research.
Moreover, we analyze the 2019 \gls{soups} proceedings, as they discuss recent research, but are mature enough that other researchers had almost five years to address suggestions from the 2019 future work statements.
Our analysis of published materials has no human subjects and contains no sensitive data, thus we do not see ethical concerns, do not further discuss ethics, and did not request IRB approval.

\subsection{Dataset}

For our systematic literature analysis, we compiled a dataset with relevant literature.
This corpus consists of two parts.

\boldparagraph{SOUPS 2019 Proceedings (n = \var{papers.soups})}
For the analysis of future work statements, we considered all \var{papers.soups}~publications from the \gls{soups} 2019 proceedings.
We collected the set of publications based on bibliographic data from \emph{DBLP}~\cite{dblp}, and validated completeness with a comparison to the \gls{soups} full proceedings and technical sessions~\cite{soups-proceedings}.

\boldparagraph{\Gls{FPS} (n = \var{papers.fps})}
To investigate the implementation of future work statements, we included all citations of the aforementioned \gls{soups} 2019 papers.
We consider publications that cite a (\gls{soups} 2019) paper potential future work of the original paper.
On the contrary, as good scientific practice means citing and crediting prior work whose future work statement one addresses, we assume that the non-existence of such a citation implies not implementing a future work statement.
To identify citations, we used \emph{Google Scholar}, which yields \var{citations.all}~citations of the \gls{soups} 2019 publications (as of December 2023).

We applied several inclusion and exclusion criteria. 
Of the initial \var{citations.all}~citations, we collected citing papers for the analysis and excluded all non-downloadable publications, e.g., due to rare cases of paywalls that our library does not provide access for,\footnote{Our institution provided us access to all major publishers relevant for \gls{usp}, including \emph{IEEE Xplore}, the \emph{ACM Digital Library}, and \emph{Springer} and \emph{Elsevier} journals and proceedings.}
dead links, and non-English websites and papers.
We considered conference and journal papers, poster abstracts, books, and theses, but excluded non-publication formats such as presentation slides.
Moreover, we removed any remaining non-English publications, publications whose PDF files we could not open, and duplicate citations per \gls{soups} 2019 publication during the analysis. 
Applying all inclusion and exclusion criteria left \var{papers.fps}~papers from \var{citations.following}~citations of \gls{soups} 2019 publications for the analysis (cf.\ \autoref{fig:method-overview}).

\subsection{Data Analysis}
Below, we describe our two-phase data analysis process. 

\subsubsection{Analyzing Future Work Statements}

In the \var{papers.soups}~\gls{soups} 2019 papers, we identified and analyzed all future work statements. 
The first author read all publications~\cite{keshav2007read} (only skipping related work sections as we did not expect them to contain future work) and marked all future work statements.
We define a future work statement as follows:
\begin{description}
   \item[Future Work Statement:] A future work statement is a passage in a research article, that suggests future work ideas that the research community could address.
\end{description}
We marked all future work statements and summarized their content using the qualitative data analysis software \emph{Atlas.TI}\footnote{\url{https://atlasti.com/}}.
For all identified future work statements, we coded content, location, and specificity. 
The categories used for content and specificity are presented in Tables~\ref{tab:fws-categories} and \ref{tab:specificity}, respectively.

\subsubsection{Analyzing \gls{FPS}}

For the \gls{fps}, i.e., any of the \var{papers.fps}~publications citing the \gls{soups} 2019 papers, we analyzed if and how any of the initially identified future work statements were implemented.
To assess whether citing publications implement future work of the initial SOUPS papers, the first author reviewed all citations for each \gls{soups} 2019 paper.
For each citing paper, the researcher read its title and abstract and examined the context---location and level of detail---of citing the initial \gls{soups} 2019 paper to identify whether the paper implements any of its future work.
This encompassed instances where the authors directly stated that they implement suggestions from a future work statement, and instances where a paper's contributions match the content of a future work statement.
As we assumed that most citing papers do not implement future work statements of a \gls{soups} 2019 publication, we based the assessment on the title, abstract, and the citation itself and only considered other sections in cases of doubt, similar to the three pass approach for reading papers~\cite{keshav2007read}.
For each paper that implemented future work, we performed an in-depth analysis and focused on how exactly the future work statement of the initial \gls{soups} 2019 paper was implemented, if the future work statement was acknowledged, or if it was just general follow-up work (e.g., reusing methods, replications). 
After carefully analyzing a \gls{soups} 2019 publication and all its citations, we continued with the next \gls{soups} 2019 paper, until we finished all \gls{soups} 2019 papers with future work statements and \var{papers.fps}~publications.

\subsection{Limitations}
We note that this work has some limitations that are typical for systematic literature analyses.
First, we used citations to identify the papers that potentially implement future work statements of prior publications.
This assumes that the scientific community properly cites scientific work that is related to or on which it built, e.g., when implementing its future work statements. 
If not cited, our approach misses these publications.
However, we argue that the vast majority cites papers correctly, so we would only miss a negligible amount.
Moreover, we rely on Google Scholar data, which is a common source~\cite{conf/ssp/LadisaPMB23}, but not guaranteed to be fully accurate. %

Our insights are limited to one year of \gls{soups} proceedings. 
However, given the immense effort to consider all citations, we deem this an acceptable trade-off, as we are not aware of any indicators for changes in future work behavior. 
That said, focusing on the 2019 \gls{soups} proceedings allows the following papers to emerge over almost five years---while more recent proceedings would allow fewer papers that implement future work statements to be published.

Last, we acknowledge that mainly a single researcher conducted the analysis.
While multiple researchers often conduct the same analysis, we refrain from this, due to the high effort of checking \var{citations.following} citations. 
However, two people looked at all future work statements, and edge cases in the analysis of \gls{fps} were discussed with the research team to resolve conflicts and reach consensus.
Moreover, we argue that future work statements and their content should be clear recommendations that can be reliably detected and analyzed. 

\subsection{Positionality Statement}
As parts of this paper are qualitative and require some researcher interpretation, we disclose our backgrounds, which might influence results~\cite{journal/lancet/Malterud01}.

The research group consists of four researchers. 
The main author, who largely conducted the analysis, is a Master's student in computer science and is familiar with \gls{usp} through various university courses and being a co-author in another paper project.
Moreover, the team consists of two Ph.D.\ students and a full professor who are all trained computer scientists and have experience with reading, reviewing, conducting, and publishing \gls{usp} research.
They helped with writing the paper and supervised the project.

\section{Future Work Statements in \gls{soups} Publications}
\label{sec:results-rq1}

\begin{figure}
    \centering
    \includegraphics[width=\linewidth]{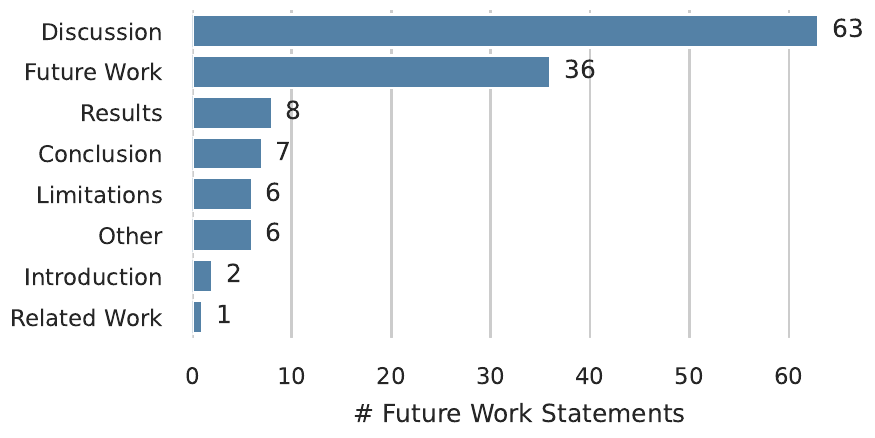}    
    \caption{Distribution of future work statements in different sections in the \gls{soups} 2019 proceedings.}
    \label{fig:location}
\end{figure}

We find that \var{papers.fws} of \var{papers.soups} papers in the 2019 \gls{soups} proceedings contain \var{fws.total} future work statements in total, with a median of \var{fws.median} future work statements per paper (SD=\var{fws.std}). 
This is slightly above the analysis of information system design papers by~\citeauthor{journal/PAJAIS/BarataC23}~\cite{journal/PAJAIS/BarataC23}.
\autoref{tab:result-future-work} illustrates the number of future work statements for each paper.
Future work statements had an average length of \var{fws.words.avg} words (MD=\var{fws.words.median}, SD=\var{fws.words.std}).
This is on average 10 words longer than \citeauthor{conf/asist/ZhuWS19} found for NLP publications~\cite{conf/asist/ZhuWS19}, which may be because we considered surrounding contextualizing sentences in our manual analysis, while their automatic approach did not.
Most future work statements were located in the discussion section followed by sections containing ``future work" in the title, which we found in eight papers. 
However, some were also placed in less obvious places such as next to specific results, where they are at a higher risk of being overlooked by readers.
Five papers advertised future work statements in their abstract, and seven additional papers declared them in a contribution summary or paper structure in the introduction.
\autoref{fig:location} provides an overview of future work statement locations.
We describe the different types of future work statements, their content, and their specificity below.

\subsection{Content of Future Work Statements}
\label{sec:results-rq1.content-of-FWS}
We assigned six categories to the future work statements based on their content, four of which we adapted from prior work~\cite{conf/asist/ZhuWS19}, and two that emerged from the data.
Multiple categories are possible for one future work statement. 
Five future work statements were too ambiguous to assign any category.\footnote{Those are marked in the replication package.}
We present an overview of the categories, their definition, and the number of corresponding future works in \autoref{tab:fws-categories}.

\begin{table*}[ht]
    \caption{Summary of the six future work statement categories we identified and their distribution in the \gls{soups} 2019 proceedings.}
    \label{tab:fws-categories}
    \centering
    \rowcolors{2}{white}{gray!10}
    \begin{threeparttable}
    \begin{tabularx}{\linewidth}{lXr}
        \toprule
         \textbf{Category} & \textbf{Definition} & \textbf{\#FWS\rlap{\tnote{1}}} \\
         \midrule
         Future Research Target\textsuperscript{$\ast$} & The future work statement points out a specific area, a topic, or a goal for future research. & 49\\
         Potential Factors of Influence\textsuperscript{$\ast$} & The future work statement asks the researcher to further investigate the effect of potentially influential factors uncovered in the study. & 34\\
         Extension of Methodology & The future work statement asks future research to address limitations inherent to the study's methodology or to verify or expand the results with an extended or different methodology. & 24\\
         Supportive\textsuperscript{$\ast$} & The future work statement describes how the contributions of the study can support future research design decisions. & 12\\
         Different Populations & The future work statement calls for an examination of (a) different population(s). & 11\\
         Utilize Contributed Methodology\textsuperscript{$\ast$} & The future work statement calls for future research to utilize a methodology or study instrument that the study contributes. & 5\\
         \bottomrule
    \end{tabularx}
    \begin{tablenotes}
        \item [$\ast$] Adapted from \citeauthor{conf/asist/ZhuWS19}~\cite{conf/asist/ZhuWS19}.
        \item [1] Number of future work statements.
        \end{tablenotes}
    \end{threeparttable} 
\end{table*}

\boldparagraph{Present Future Research Targets}
Most commonly, 49 future work statements described a target for future work, such as a new area to investigate, or a new research goal.
These were often related to the study's topic but did not necessarily follow directly from its contributions. 
Instead, there were often other interesting aspects of the research subject that had been out of the original study's scope, such as:
\blockquote[{\cite{conf/soups/HabibZJSSACSS19}}]{Our analysis primarily focuses on usability issues and does not intend to analyze legal compliance (although the latter is an important direction for future work).},
and
\blockquote[{\cite{conf/soups/DruryM19}}]{We also intend to further explore the question of how robust the subject Organization is against active attacks, and if subject spoofing might become more common in the future.}.
In addition, several of these future work statements proposed research goals to help reduce or solve problems that a study had uncovered, e.g.: \blockquote[\cite{conf/soups/LiRMMC19}]{Research into the usability of dynamic updating systems and avenues for improvement could potentially eliminate update timing concerns for administrators in the future.}.

\boldparagraph{Identify Potential Factors to Investigate}
34 of the future work statements encouraged future research to investigate potential factors that could influence a phenomenon the study had observed. 
Often, these potential factors directly emerged from the study's findings, but insights regarding their real effect were limited and warranted further investigation.
Qualitative methodology uncovered potential factors to investigate, such as in this study investigating security and privacy attitudes of older adults:
\blockquote[\cite{conf/soups/FrikNBLSE19}]{Some of the patterns we identified in our exploratory qualitative study merit further systematic investigation, such as older adults’ uncertainties about data deletion and retention}.
Some studies had results on potential factors that were not their primary research target, and inconclusive due to confounding factors:
\blockquote[\cite{conf/soups/DasDH19}]{Additional research will be necessary to tease apart the effect of culture from other confounding factors such as,  for example, the work contexts of AMT workers in the U.S.
versus those in India.}.
Some studies also reasoned about potential factors beyond their results:
\blockquote[\cite{conf/soups/FrikNBLSE19}]{Finally, the measures we recommend should be tested “in the wild” to determine their efficacy. For example, we might test whether having targeted training materials for educational programs can positively impact older adults’ privacy and security behaviors}.

\boldparagraph{Call for Extension of the Methodology}
In 24 cases, future work statements addressed a study's methodological limitations and called for future work to overcome them with an extended or different methodology.
One example of such a limitation was bias introduced by the method of data collection:
\blockquote[\cite{conf/soups/DasDH19}]{In future work, it would be useful to catch behavior changes closer to the moment those behaviors occur, perhaps through a diary study.}
Several future work statements also asked for confirmation of the study's results or their generalization, e.g.,
\blockquote[\cite{conf/soups/VanceEJKA19}]{Second, our experiment was designed to expose participants to notifications at a higher rate than is normally encountered in the same amount of time during usual computer usage. In future research, it would be interesting to explore if the generalization of habituation occurs with the same amount of exposures distributed across a longer time window.}
One study that contributed a survey instrument pointed out a need for further and continued evaluation of the scale~\cite{conf/soups/FaklarisDH19}.

\boldparagraph{Detail How Results Support Future Research Design Decisions}
Twelve future work statements included a description of how the study's contributions benefit future work in the design of studies. 
This was often applied to rather broad areas of research.
Supportive future work statements commonly revolved around methodological design decisions, including ethical considerations as well as insights on characteristics of populations to be considered in study design:
\blockquote[\cite{conf/soups/HayesKPW19}]{However, privacy designs need to consider the multi-faceted and intersectional aspects of people’s marginalized identities. We believe that one practical way to do this is through focus groups and participatory action/design research.}
Some studies instead highlighted how their results could inform the design of systems and solutions in a particular research area:
\blockquote[\cite{conf/soups/MeckeBKPA19}]{For example, work on systems for mimicry attacks could use our results here as a baseline for unsupported modification ability.}.

\boldparagraph{Call to Examine Different Populations}
Future work statements mentioned studying one or more different populations eleven times.
Most common was a desire to extend the studied population to allow for generalization of results or gain insights on additional groups, such as:
\blockquote[\cite{conf/soups/WuGHTVZS19}]{It would be helpful to study populations with different risk-cost trade-offs, such as immigrants or dissidents, and to ascertain that risk communication translates well to other cultures and languages.}.
Two future work statements highlighted the value of comparing results from different populations, and one said that research focusing on a specific population may enhance their general results.

\boldparagraph{Call for Studies Utilizing a Contributed Methodology}
The five instances of future work statements recommending further utilization of contributed methods occurred across four papers. 
One contributed a method to evaluate privacy designs and highlighted the possibility of using it across systems and scenarios.
One's main contribution was the development of a scale for assessing security attitudes and listing potential research questions it could help answer as examples. 
The third presents a system design to simulate an implicit authentication system used in the main study and expresses hope to enable further evaluations.
Finally, one paper evaluated some usability principles and concluded that their approach was suitable to also evaluate other such principles.

\subsection{Specificity of Future Work Statements}
\begin{figure}[ht]
    \centering
    \includegraphics[width=\linewidth]{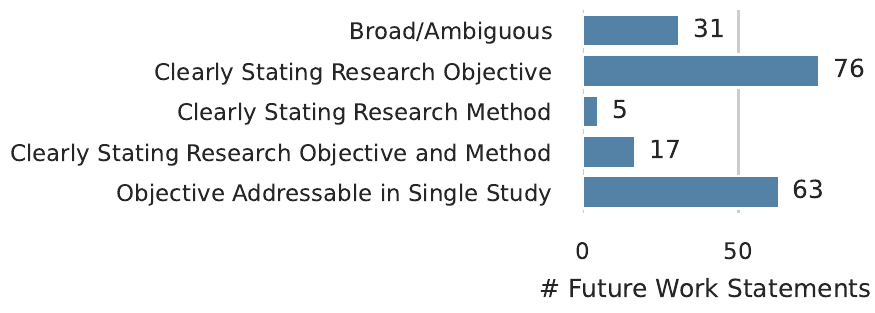}
    \caption{Summary of future work statement specificity.}
    \label{tab:specificity}
\end{figure}

In addition to the type of content, we analyzed the level of specificity of future work statements. 
We identified four different levels: broad/ambiguous, clearly stating a research objective, clearly stating a research method, and clearly stating a research objective and method. 
For future work statements that stated a research objective, we estimated if the objective could be addressed in a single study. 
We present an overview of how many future work statements were assigned to what level in \autoref{tab:specificity}.

We note that while many future work statements provided a research objective, 31 of them were too broad to be addressed by a single follow-up study, and only a few commented on possible methods to achieve the research objective. 
Several research objectives were very ambitious, and might not be achievable with methods currently available to \gls{usp} research, e.g.:
\blockquote[\cite{conf/soups/MhaidliZS19}]{explore new monetization models  for apps that go beyond advertising and paid models, which  ideally retain the low barrier to entry that free apps have, but  do not pose the same consumer risks as targeted advertising.}, or
\blockquote[\cite{conf/soups/KumRIRLS19}]{more research is needed to design optimal systems to induce best behavior.}.
At 31, about a quarter of the future work statements were broad or ambiguous and provided neither a research objective nor method.
While 121 future work statements did not specify who should implement them beyond the research community, eight statements were authors declaring their intent to conduct the research they suggested in their future work statements. 
All of these presented a research objective with a scope suitable for a single study.

\begin{summaryBox}{Key Findings (RQ1)}
    \ We found \var{fws.total}~future work statements that commonly point out new research areas, topics, and goal. 
    They also called for the investigation of the impact of potential factors of influence, methodological extensions, consideration of insights in future research design, examination of additional populations, and reuse of methodological contributions.
    Future work statements were often broad or ambiguous, or meant to address the limitations of a paper. 
\end{summaryBox}

\section{Implementation of Future Work Statements}
\label{sec:results-rq2}
Below, we detail our findings on how \gls{fps} implement suggestions from future work statements of the \gls{soups} 2019 proceedings.
\autoref{tab:result-future-work} gives an overview of the \gls{fps} that implement and acknowledge future work statements.

\newcommand{\citeandyear}[1]{\citeauthor{#1}~\cite{#1}~(\citeyear{#1})}

\begin{table*}[tb]
    \centering
    \begin{threeparttable}
        \caption{Overview of all \var{papers.soups} \gls{soups} 2019 publications, their future work statements, and \gls{fps} which implemented the future work statements.}
        \label{tab:result-future-work}
        \small
        \rowcolors{2}{white}{gray!10}
        \begin{tabularx}{\linewidth}{r>{\hspace{-8pt}}lr>{\hspace{10pt}}cr>{\hspace{12pt}}crX}
            \toprule
                \multicolumn{2}{l}{\textbf{SOUPS 2019 Proceedings}} &
                \textbf{\#Cites\rlap{\tnote{1}}} &
                \textbf{\makecell{Has\\FWS\rlap{\tnote{2}}}} &
                \textbf{\#FWS\rlap{\tnote{2}}} &
                \multicolumn{3}{l}{\textbf{Implementation and Acknowledgment of Future Work Statements}} \\
            \midrule
                \cite{conf/soups/AbdiRS19} & \citeauthor{conf/soups/AbdiRS19} & 2 & \yes & 4 & \no & 0 & --- \\
                \cite{conf/soups/AlqhataniL19} & \citeauthor{conf/soups/AlqhataniL19} & 44 &  \yes & 2 & \no & 0 & --- \\
                \cite{conf/soups/AyalonT19} & \citeauthor{conf/soups/AyalonT19} & 13 & \yes & 7 & \no & 0 & --- \\
                \cite{conf/soups/BusseSS19} & \citeauthor{conf/soups/BusseSS19} & 54 & \yes & 4 & \no & 0 & --- \\
                \cite{conf/soups/CiolinoPD19} & \citeauthor{conf/soups/CiolinoPD19} & 51 & \no & 0 & \no & 0 & --- \\
                \cite{conf/soups/DasDH19} & \citeauthor{conf/soups/DasDH19} & 51 & \yes & 4 & \yes & 1 & \citeandyear{murthy2021individually} \\
                \cite{conf/soups/DiMartinoRWQLA19} & \citeauthor{conf/soups/DiMartinoRWQLA19} & 62 & \yes & 6 & \no & 0 & --- \\
                \cite{conf/soups/DruryM19} & \citeauthor{conf/soups/DruryM19} & 41 & \yes & 4 & \no & 0 & --- \\
                \cite{conf/soups/FaklarisDH19} & \citeauthor{conf/soups/FaklarisDH19} & 66 & \yes & 7 & \no & 0 & --- \\
                \cite{conf/soups/FrikNBLSE19} & \citeauthor{conf/soups/FrikNBLSE19} & 135 & \yes & 7 & \yes & 1 & \citeandyear{ray2020warn} \\
                \cite{conf/soups/FultonGMARM19} & \citeauthor{conf/soups/FultonGMARM19} & 37 & \yes & 4 & \no & 0 & --- \\
                \cite{conf/soups/HabibZJSSACSS19} & \citeauthor{conf/soups/HabibZJSSACSS19} & 81 & \yes & 2 & \no & 0 & --- \\
                \cite{conf/soups/HayesKPW19} & \citeauthor{conf/soups/HayesKPW19} & 50 & \yes & 8 & \no & 0 & --- \\
                \cite{conf/soups/KumRIRLS19} & \citeauthor{conf/soups/KumRIRLS19} & 16 & \yes & 10 & \no & 0 & --- \\
                \cite{conf/soups/LiRMMC19} & \citeauthor{conf/soups/LiRMMC19} & 60 & \yes & 7 & \yes & 1 & \citeandyear{martius2020does} \\
                \cite{conf/soups/MeckeBKPA19} & \citeauthor{conf/soups/MeckeBKPA19} & 12 & \yes & 11 & \no & 0 & --- \\
                \cite{conf/soups/MeckeRBPA19} & \citeauthor{conf/soups/MeckeRBPA19} & 4 & \yes & 5 & \no & 0 & --- \\
                \cite{conf/soups/MhaidliZS19} & \citeauthor{conf/soups/MhaidliZS19} & 64 & \yes & 7 & \no & 0 & --- \\
                \cite{conf/soups/PatnaikHR19} & \citeauthor{conf/soups/PatnaikHR19} & 50 & \yes & 3 & \no & 0 & --- \\
                \cite{conf/soups/PearmanZBCC19} & \citeauthor{conf/soups/PearmanZBCC19} & 141 & \yes & 5 & \no & 0 & --- \\
                \cite{conf/soups/QinLJFAGRV19} & \citeauthor{conf/soups/QinLJFAGRV19} & 27 & \yes & 2 & \no & 0 & --- \\
                \cite{conf/soups/ReeseSDACS19} & \citeauthor{conf/soups/ReeseSDACS19} & 160 & \yes & 2 & \yes & 2 & \citeandyear{kruzikova2022usable}, \citeandyear{PaperID378} \\
                \cite{conf/soups/SimoiuBGG19} & \citeauthor{conf/soups/SimoiuBGG19} & 41 & \yes & 7 & \no & 0 & --- \\
                \cite{conf/soups/TabassumKL19} & \citeauthor{conf/soups/TabassumKL19} & 136 & \yes & 5 & \yes & 2 &  \citeandyear{sereda2022supporting}, \citeandyear{haney2020smart} \\
                \cite{conf/soups/VanceEJKA19} & \citeauthor{conf/soups/VanceEJKA19} & 36 & \yes & 3 & \no & 0 & --- \\
                \cite{conf/soups/VoronkovML19} & \citeauthor{conf/soups/VoronkovML19} & 21 & \yes & 1 & \no & 0 & --- \\
                \cite{conf/soups/WuGHTVZS19} & \citeauthor{conf/soups/WuGHTVZS19} & 29 & \yes & 2 & \no & 0 & --- \\
            \midrule
            \multicolumn{2}{l}{\textbf{Total}} & \var{citations.all} & \var{papers.fws} & \var{fws.total} & \var{papers.fw.implemented} & \var{papers.fps.implementing-fws} & \\
        	\bottomrule
        \end{tabularx}
        \begin{tablenotes}
        \item [1] Number of citations of the SOUPS 2019 paper (as of December 2023).
        \item [2] Number of future work statements in the SOUPS 2019 paper.
        \end{tablenotes}
    \end{threeparttable}%
\end{table*}

\subsection{Implementation Frequency of Future Work Statements}
\label{sec:results-rq2.implementation-frequency-of-FWS}

Out of the \var{papers.fws}~\gls{soups} 2019 publications that contain future work statements, for \var{papers.fw.implemented.text}~of them suggestions from future work statements have been implemented and acknowledged by \var{papers.fps.implementing-fws.text}~\gls{fps}.
This is similar for the \var{fws.total}~future work statements from the \var{papers.fws}~\gls{soups} 2019 publications; \var{fws.implemented.text}~statements have been implemented and acknowledged.
While future work statements of \var{papers.amount.soups-with-one-future-paper} \gls{soups} 2019 publications have been implemented by one \gls{fp}, the remaining \var{papers.amount.soups-with-two-future-paper.text} had their future work statements implemented by two \gls{fps}.

Overall, we see a slight tendency for more future work implementations among highly cited \gls{soups} 2019 publications, compared to the ones that are less often cited.
While all papers whose future work statements have been implemented by two \gls{fps} have at least 136~citations, those with only one \gls{fp} implementing a future work statement have only been cited at least 51~times (cf.\ \autoref{tab:result-future-work}).%
\footnote{\textbf{Disclaimer:} This is only a hypothesis based on our findings, as the number of papers in the \gls{soups} 2019 proceedings is too small to statistically confirm this observation.}

\begin{figure}[ht]
    \centering
    \includegraphics{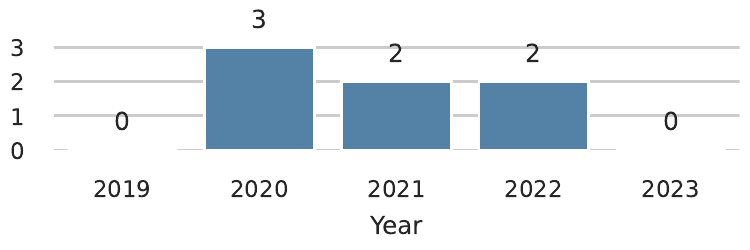}
    \caption{Publication years of \gls{fps} that implement a future work statement from \gls{soups} 2019 papers.}
    \label{tab:years-fws-implementation-published}
\end{figure}

As shown in \autoref{tab:years-fws-implementation-published}, the future work statements have been first implemented in the year following the publication of the original \gls{soups} paper.
While we found a similar occurrence in the following years, in 2023 no \gls{fps} implemented any future work statement from a \gls{soups} 2019 paper.

\subsection{How Future Work Statements were Implemented}
\label{sec:results-rq2.how-fws-were-implemented}

Analyzing the \gls{fps} that cite \gls{soups} 2019 papers, we found \var{papers.fps.implementing-fws.text}~publications that implement and acknowledge the suggestions from a future work statement. 
However, we found five other types of related follow-ups that are similar to but not implemented and acknowledged future work statements. 
Below, we outline the qualitative differences.

\boldparagraph{Implementation of Future Work Statements}
We found \var{papers.fps.implementing-fws.text}~publications (\autoref{tab:result-future-work}) that implement and acknowledge future work statements as outlined in the original \gls{soups} 2019 publication.
By implementing future work statements, we refer to follow the suggestions the authors of the preceding \gls{soups} paper made exactly or with only minor differences.
By acknowledgment, we mean that the authors of the \gls{fp} state that they were aware of the implemented future work statements from prior publications.

For example, \citeauthor{conf/soups/ReeseSDACS19} published a SOUPS article named \citetitle{conf/soups/ReeseSDACS19}, containing a future work statement which recommends:
\blockquote[\cite{conf/soups/ReeseSDACS19}]{Even if we had reached saturation, the limited demographics of the study still warrant further studies with a broader population.}.
\citeauthor{kruzikova2022usable} implemented this future work statement and acknowledged it as follows:
\blockquote[\cite{kruzikova2022usable}]{However, there are few studies that focus on these underlying factors, and they were often conducted on small or student samples that limit the generalization to the wider population (e.g., Weir et al., 2009;  Krol et al., 2015; Reese et al., 2019). To overcome this limitation, we collected data from two independent samples of smartphone users that together cover users aged 26 to 82.}.

In another paper, the authors acknowledged and implemented the entirety of future work statements, without implementing each exactly. 
\citeauthor{conf/soups/DasDH19}~\cite{conf/soups/DasDH19} identified differences in how people from the U.S.\ and India share \gls{sp} behavior, and called to investigate this further in three future work statements:
They called to investigate
(i)~the effect of culture and other confounding factors between workers from India and the U.S.,
(ii)~the level of age-based personalization needed to trigger \gls{sp} behavior.
They also stated:
\blockquote[{\cite{conf/soups/DasDH19}}]{People from the U.S.\ were far less likely to share than people from India. \textelp{}, but further research is necessary for this to be conclusive.}.
In a \gls{fp}, \citeauthor{murthy2021individually}~\cite{murthy2021individually} addressed and acknowledged these future work statements as follows:
\blockquote[\cite{murthy2021individually}]{Prior work comparing the privacy attitudes and behaviors of people in the US and India suggests that people in India take a more social approach to SP [10].}.
They implemented future work by investigating an older adult population in India and their social \gls{sp} practices and social support system regarding \gls{sp}.

Given the high number of \var{papers.fps}~analyzed publications that cite a \gls{soups} 2019 paper, the implementation and acknowledgment of future work statements were rare.
Instead, the below types---which are similar to, but not implementation of future work statements, as discussed before---occured more frequently.

\smallskip
\bigskip

Below, we outline the five types of related follow-up studies that are similar to, but not implementing or acknowledging, a future work statement.
We give examples and explain why we do not consider these to implement and acknowledge future work statements.
As these insights are qualitative, we refrain from reporting numbers on their occurrence and only roughly indicate prevalence.

\boldparagraph{Thematic Similarity with Future Work Statements}
\label{results:thematic-similarity-with-fws}
Most commonly, we found publications that did align with future work statements to some extent but lacked a clear acknowledgment of the future work statements in the original papers.
We call this \emph{thematic similarity} as the \gls{fp} might be inspired by the prior work and/or its future work statements.
There are several examples of future work statements that \gls{fps} address to some degree, such as:
\blockquote[{\cite{conf/soups/FrikNBLSE19}}]{In particular, it is not yet clear how the issues we identified affect older adults’ privacy and security behavior as compared to the general population, or whether their security and privacy management strategies are more or less effective than those of the general population.}
While the paper by \citeauthor{berridge2022control}~\cite{berridge2022control} implement but do not acknowledge future work statements of that paper~\cite{conf/soups/FrikNBLSE19}, others' publications were related but aligned only roughly with the future work statements.
For example, one paper~\cite{PaperID1200} conducted a related study and compared privacy attitudes, trust, and risk beliefs between younger and older adults;
and another paper~\cite{PaperID1202} compared the privacy decision-making processes of older and younger adults.
Another example is the future work statement by \citeauthor{conf/soups/DruryM19}:
\blockquote[{\cite{conf/soups/DruryM19}}]{In future work, we plan to explore whether the observed differences between benign and phishing website certificates can be used to enhance the phishing detection capabilities of automated detection tools or users themselves.}
Several publications address this problem as well \cite{PaperID541, PaperID544}. %
They focus on this topic but do not directly acknowledge that they were inspired by previous future work statements.
Since they did not acknowledge the future work statement from the original \gls{soups} paper---although citing it---we classified those as similar but not an implementation of future work statements.
The main argument is that it is not clear whether the authors were guided or tried to implement the future work statement.

\boldparagraph{Replication Papers}
\label{results:replication-papers}
A special case that we encountered was papers that conducted a replication of a preceding publication.
While replication papers copy or adapt the method of the original paper, we did not classify those as implementation of future work, when there was no future work statement on replication.
Nonetheless, we acknowledge that replication studies are, to some degree, a continuation of prior work and therefore similar to future work implementations.
There are two types of replication papers:
(i)~replication papers who did the replication as suggested in a future work statement,
and 
(ii)~conducting a replication without being suggested in a future work statement.
We found two replication papers~\cite{teradareplication, baigreplication} which are instances of the first type, as they loosely address future work statements without acknowledging them. 
Their authors did not relate to future work statements, therefore we do not know if they considered them.
We note that some papers suggest replications implicitly in their future work statements, e.g., by considering a different or broader population.
We found four replication papers of the second type~\cite{PaperID706,PaperID711,PaperID355,PaperID631}, that were conducted although the preceding \gls{soups} 2019 publications did not contain a future work statement that proposes a replication.

\boldparagraph{Simultaneous Publication of the Same or Similar Studies}
In a rare case, two related papers were very similar to each other.
However, they did not seem to have influenced each other as the researchers worked independently and in parallel on their projects. 
In that sense, the researchers worked on the same projects unintentionally.
Therefore, they did not implement future work statements of each other.
\citeauthor{conf/soups/TiefenauHKZ20} mentioned in \citetitle{conf/soups/TiefenauHKZ20}~\cite{conf/soups/TiefenauHKZ20} that \citeauthor{conf/soups/LiRMMC19}~\cite{conf/soups/LiRMMC19} published independently a \enquote{closely related paper} at \gls{soups} 2019. 
Hence, it is not the implementation of a future work statement, but we acknowledge their strong connection.

\boldparagraph{Methodical Contribution and Reuse}
One of the \gls{soups} 2019 papers is a methodological contribution.
\citeauthor{conf/soups/FaklarisDH19} presented \emph{SA-6}, a scale to assess security attitudes of human subjects~\cite{conf/soups/FaklarisDH19}.
There are multiple \gls{fps} which use this scale in their studies~\cite{10.1145/3411764.3445122, conf/usenix/AbrokwaDAM21, conf/soups/MayerMMA22}. %
Therefore, there is a clear relation to the original \gls{soups} paper, but not in terms of implementing future work statements.
In another case, the authors reused and adapted methods from \citeauthor{conf/soups/MhaidliZS19}~\cite{conf/soups/MhaidliZS19}:
\blockquote[\cite{PaperID181}]{Our survey was inspired by the work of Mhaidli et al.~[183] and consisted of five parts.}.

\boldparagraph{General Inspiration without a Future Work Statement}
Last, we found \gls{fps} that generally seem to be highly related to, building on, or extending prior publications, even if prior publications do not contain any future work statement.
We perceived this as a general advancement of science to inspire new ideas from related research (even in the absence of future work statements); but it is no implementation of future work statements.
As previously described, there was initial work at \gls{soups} from \citeauthor{conf/soups/DruryM19}~\cite{conf/soups/DruryM19}.
Three years later, \citeauthor{PaperID538} published \citetitle{PaperID538}~\cite{PaperID538} which seems to be highly related and inspired by the publication of \citeauthor{conf/soups/DruryM19}:
\blockquote[{\cite{PaperID538}}]{Previous work [57] suggests that it is generally impossible to differentiate between benign sites and phishing sites based on the content of simple X.509 certificate features alone. To address this issue, we combine aggregate and historical certificate features taken from CT logs to effectively identify recurring long-term phishing domains. We also combine CT and pDNS features to effectively mark new phishing domains.}. %

\subsection{Who Implements Future Work Statements}
\label{subsec:who-implements}

For the \var{papers.fps.implementing-fws.text}~\gls{fps} that implement and acknowledge future work statements, we investigated who implemented them and how those papers have been published.

\boldparagraph{Authors}
Only \var{fps.implemented.author.text} \gls{fps} that implements \var{fws.implemented.author.text} future work statements share at least one author with the original \gls{soups} 2019 paper.
For the remaining \var{fws.implemented.others.text}, a distinct set of authors implemented the future work statement.

\boldparagraph{Venues \& Publication Types}
Overall, the \var{papers.fps.implementing-fws.text} \gls{fps} that implement future work statements were different types of publications, and published across different venues.
At \var{fps.implemented.conferences.text}, most were full conference papers published at \gls{cscw}, \gls{pets}, \gls{usenix}, and \gls{hci-cpt}.
None were also published at \gls{soups}.
The remaining publications were a journal article in Computer\&Security, a workshop paper, and a thesis.

\subsection{Characteristics of Implemented Future Work Statements}
\label{subsec:fws-char}
Of the \var{fws.implemented.text} future work statements that were implemented and acknowledged, none declared the authors' intent to conduct future work, which aligns with the fact that most were implemented by a distinct set of authors (cf.\ \autoref{subsec:who-implements}).
The statements' content varied from method extensions~(4), to potential influence factors~(3), analysis of different populations~(3), and statements supporting future research design~(2). 
None suggested a future research target.
Six of the future work statements stated a research objective, two were broad or ambiguous, and none suggested a research method for future work.

\begin{summaryBox}{Key Findings (RQ2)}
    \ We find that \var{fws.implemented.text} of \var{fws.total} future work statements have been implemented and acknowledged by \var{papers.fps.implementing-fws.text}~\gls{fps}, starting the year after the original \gls{soups} publication.
    Publications inspired future research (e.g., thematic similarity, method reuse, replications) without acknowledgment of future work statements.
    Authors rarely implemented their own future work statements.
    Implemented and acknowledged statements often called for an extension of their methodology. 
\end{summaryBox}

\section{Discussion}
\label{sec:discussion}

In our analysis of the \var{papers.soups} \gls{soups} 2019 publications and their future work statements \emph{(RQ1)}, we find that almost all papers (\var{papers.fws}) include future work statements, \var{fws.total} in total (\autoref{sec:results-rq1}).
However, we find that many future work statements are broad, ambiguous, or have a large scope (\autoref{sec:results-rq1.content-of-FWS}).
Regarding how researchers implement those future work statements \emph{(RQ2)}, we find that only \var{fws.implemented.text} have been implemented and acknowledged by \var{papers.fps.implementing-fws.text}~publications (\autoref{sec:results-rq2.implementation-frequency-of-FWS}).
Nonetheless, publications might inspire future and follow-up work---independent of whether any future work statement was included (\autoref{sec:results-rq2.how-fws-were-implemented}). 

Given the low number of implemented and acknowledged future work statements, we conclude that the overall impact of current future work statements on the \gls{usp} field and its advancement is limited. 
We hypothesize that the high prevalence of broad and ambiguous future work statements is one reason for the limited impacts, as such statements are rarely actionable.
Additionally, we find that future research might be directed by prior, related research papers as a whole---not only their future work statements.
Method reuse and thematic similarity to papers and their future work statements (\autoref{sec:results-rq2.how-fws-were-implemented}) suggest that future research might be more dependent on prior publications' results, innovative ideas, methods, and broader context instead.
Replication studies, while rare, indicate this as well, as they are also conducted when not proposed by future work statements.

While we do not know the authors' motivation for specifying future work statements in their papers, we see a few potential reasons.
As authors might want to keep good research ideas for their research agenda, but still want to relate to a prior publication, they might include only broad and unspecific statements to protect themselves from being scooped.
Some authors stated that they would implement the suggestions from their future work statement in a follow-up study.
We hypothesize that this could be intended as a reservation to prevent other researchers from working on the same project.
Finally, and similar to related work~\cite{journal/PAJAIS/BarataC23}, we find that future work statements often address methodological limitations. 
Such future work statements commonly asked future research to extend methodology (cf. \autoref{sec:results-rq1.content-of-FWS}) to eliminate biases or assert generalizability of findings.
The reason for such future work statements may be to placate critical reviewers and alleviate concerns about the impact of findings rather than to actually encourage future studies.

Given the rare adoption and limited impact of future work statements in our dataset, the question arises whether authors should include future work statements in their publications.
If future work is only stated to refute limitations or is too broad to be actionable, we argue that future work statements add little value to a research publication.
In some cases, 
future work opportunities are trivial.
For example, replication is an obvious follow-up idea for almost any paper.
If a future work statement adds little value, authors might better use space for more valuable things.
Moreover, future work statements should not be included because they are common in \gls{usp} papers.
Despite all that, researchers spend a lot of time on their research projects and might have unique insights on the topic, including possible future work, that are not evident to readers of their papers.
In such cases and generally, when they add novel or valuable points to a paper, we advocate including future work statements in publications.
Below, we detail how we think of giving and implementing future work statements in \gls{usp} research can be improved.

\subsection{How to Improve Future Work Statements}
\label{sec:discussion-recommendations}
From our insights on future work statements, their content, and impact, we derive the following recommendations for the \gls{usp} community:

\boldparagraph{If Making a Future Work Statement, Be Specific}
If authors decide to include one or more future work statements in their publications, they should make them specific and actionable to help maximize value and impact. 
We observed that future work statements that were acknowledged and implemented mostly stated clear research objectives, which commonly described how to extend the study's method or population or examine uncovered potential influence factors (\autoref{subsec:fws-char}).
We believe that the \gls{usp} community can best benefit from future work statements that are based on the researchers' detailed insights into the used methods and studied problem.
Consequently, authors should take the time and space to derive future work statements from such insights and provide details on research objectives and potential methods.
Similar to \citeauthor{journal/PAJAIS/BarataC23}~\cite{journal/PAJAIS/BarataC23}, we encourage authors to include examples of possible follow-up studies and clearly explain the capabilities of datasets and artifacts.
Additionally, future work statements that provide parameter sets (e.g., what population) for replication could simplify and encourage replication studies, of which we observed relatively few in our dataset.

\boldparagraph{Make Future Work Statements Findable}%
As visible in the widespread distribution of future work statements across different paper sections (\autoref{fig:location}), there is currently no consensus in the \gls{usp} community on where to provide future work statements, making some of them hard to find.
To increase accessibility, we recommend that researchers place future work statements in the discussion section, which is currently the most common, and ideally use a designated subsection title.
We suggest to incorporate this recommendation in future guides for reporting in \gls{usp} papers, such as previously provided by \citeauthor{how-to-soups}~\cite{how-to-soups} and \citeauthor{journal/tochi/DistlerFHKLLCK21}~\cite{journal/tochi/DistlerFHKLLCK21}.

\boldparagraph{Acknowledge When Research was Shaped by Future Work Statements}
We observed that subsequent studies rarely acknowledged future work statements even when they are thematically (very) similar to them. 
This remained true for authors who followed their own future work statements and made it difficult to assess if a future work statement influenced a later study design.
To encourage \gls{usp} researchers to invest time and effort into specific, actionable future work statements, they should be explicitly acknowledged when implemented by future work.
This provides both recognition for the original authors and improved clarity on how studies build upon each other in the \gls{usp} field.

\subsection{Future Work}
In writing a paper on future work, we aim to give future work statements that follow the aforementioned recommendations.
While this paper comprehensively investigates the future work statements in \gls{soups} 2019 papers and how they are implemented, we lack insights for other years and venues.
Therefore, we suggest to conduct similar studies for other years and venues, e.g., to validate our results.
To this end, our methodology and coding could be directly replicated by other researchers.
As outlined in \autoref{sec:related-work}, several \gls{nlp}-based approaches for future work analyses exist. 
We propose applying those to facilitate a literature review of future work statements on a larger corpus of \gls{usp} publications. 
Our provided dataset of future work statements can be used as ground truth to fine-tune models for \gls{usp} research or evaluate the existing NLP approaches' performance.

\section{Conclusion}
\label{sec:conclusion}

In a literature review, we analyzed the future work statements from all \var{papers.soups}~papers of the \gls{soups} 2019 proceedings.
Then, we analyzed \var{papers.fps}~papers which cited the \gls{soups} 2019 proceedings for if and how they implement these future work statements.
\var{papers.fws}~of the \gls{soups} 2019 papers contained a total of \var{fws.total}~future work statements, of which only \var{fws.implemented.text}~statements had been implemented and acknowledged by \var{papers.fps.implementing-fws.text}~\gls{fps}.
In contrast, several studies implement future work statements, or publish thematically similar studies or replications, without acknowledging the future work statements.
This mismatch indicates a very limited impact of future work statements in \gls{usp}.
Reasons for the limited impact might be that many future work statements were unspecific and broad, that some were placed in unexpected parts of a paper, or that research is inspired by a publication’s results, innovative ideas, methods, and broader
context instead.
Therefore, we recommend considering whether adding future work statements adds value to a publication, and if so making it specific, actionable, and findable.
To increase recognition for authors' efforts and clarity about research publications, the influence of future work statements should be explicitly acknowledged.

\section*{Availability}\label{sec:availability} %

To support transparency, verifiability, and repeatability of our work, we provide the following artifacts:
(1)~annotated dataset of future work statements from \gls{soups} 2019 proceedings,
(2)~dataset of \var{papers.fps} publications that cite one or more \gls{soups} 2019 papers.
The artifacts are available at: \url{https://doi.org/10.17605/OSF.IO/QKFNB}.

\newpage

\printbibliography

@inproceedings{conf/usenix/HasegawaIA24,
    author = {Ayako A. Hasegawa and Daisuke Inoue and Mitsuaki Akiyama},
    title = {How WEIRD is Usable Privacy and Security Research?},
    booktitle = {Proc.\ 33rd USENIX Security Symposium, August 14–16, 2024, Philadelphia, PA, USA},
    publisher = {USENIX},
    year = 2024
}

@article{journal/tochi/DistlerFHKLLCK21,
    author = {Distler, Verena and Fassl, Matthias and Habib, Hana and Krombholz, Katharina and Lenzini, Gabriele and Lallemand, Carine and Cranor, Lorrie Faith and Koenig, Vincent},
    title = {A Systematic Literature Review of Empirical Methods and Risk Representation in Usable Privacy and Security Research},
    year = {2021},
    issue_date = {December 2021},
    publisher = {ACM},
    volume = {28},
    number = {6},
    journal = {ACM Trans. Comput.-Hum. Interact.},
    month = {12},
    articleno = {43},
    numpages = {50},
}

@www{how-to-soups,
    author = {Stuart Schechter},
    title = {Common Pitfalls in Writing about Security and Privacy Human Subjects Experiments, and How to Avoid Them},
    year = {2010},
    url = {https://cups.cs.cmu.edu/soups/2010/howtosoups.pdf}
}

@inproceedings {conf/soups/AbdiRS19,
    author = {Noura Abdi and Kopo M. Ramokapane and Jose M. Such},
    title = {More than Smart Speakers: Security and Privacy Perceptions of Smart Home Personal Assistants},
    booktitle = {Fifteenth Symposium on Usable Privacy and Security (SOUPS 2019)},
    year = {2019},
    isbn = {978-1-939133-05-2},
    address = {Santa Clara, CA},
    pages = {451--466},
    url = {https://www.usenix.org/conference/soups2019/presentation/abdi},
    publisher = {USENIX Association},
    month = aug
}

@inproceedings {conf/soups/AlqhataniL19,
    author = {Abdulmajeed Alqhatani and Heather Richter Lipford},
    title = {{{\textquotedblleft}There} is nothing that I need to keep {secret{\textquotedblright}}: Sharing Practices and Concerns of Wearable Fitness Data},
    booktitle = {Fifteenth Symposium on Usable Privacy and Security (SOUPS 2019)},
    year = {2019},
    isbn = {978-1-939133-05-2},
    address = {Santa Clara, CA},
    pages = {421---434},
    url = {https://www.usenix.org/conference/soups2019/presentation/alqhatani},
    publisher = {USENIX Association},
    month = aug
}

@inproceedings {conf/soups/AyalonT19,
    author = {Oshrat Ayalon and Eran Toch},
    title = {Evaluating {Users{\textquoteright}} Perceptions about a {System{\textquoteright}s} Privacy: Differentiating Social and Institutional Aspects},
    booktitle = {Fifteenth Symposium on Usable Privacy and Security (SOUPS 2019)},
    year = {2019},
    isbn = {978-1-939133-05-2},
    address = {Santa Clara, CA},
    pages = {41--59},
    url = {https://www.usenix.org/conference/soups2019/presentation/ayalon},
    publisher = {USENIX Association},
    month = aug
}

@inproceedings {conf/soups/BusseSS19,
    author = {Karoline Busse and Julia Sch{\"a}fer and Matthew Smith},
    title = {Replication: No One Can Hack My Mind Revisiting a Study on Expert and {Non-Expert} Security Practices and Advice},
    booktitle = {Fifteenth Symposium on Usable Privacy and Security (SOUPS 2019)},
    year = {2019},
    isbn = {978-1-939133-05-2},
    address = {Santa Clara, CA},
    pages = {117--136},
    url = {https://www.usenix.org/conference/soups2019/presentation/busse},
    publisher = {USENIX Association},
    month = aug
}

@inproceedings {conf/soups/CiolinoPD19,
    author = {St{\'e}phane Ciolino and Simon Parkin and Paul Dunphy},
    title = {Of Two Minds about {Two-Factor}: Understanding Everyday {FIDO} {U2F} Usability through Device Comparison and Experience Sampling},
    booktitle = {Fifteenth Symposium on Usable Privacy and Security (SOUPS 2019)},
    year = {2019},
    isbn = {978-1-939133-05-2},
    address = {Santa Clara, CA},
    pages = {339--356},
    url = {https://www.usenix.org/conference/soups2019/presentation/ciolino},
    publisher = {USENIX Association},
    month = aug
}

@inproceedings {conf/soups/DasDH19,
    author = {Sauvik Das and Laura A. Dabbish and Jason I. Hong},
    title = {A Typology of Perceived Triggers for {End-User} Security and Privacy Behaviors},
    booktitle = {Fifteenth Symposium on Usable Privacy and Security (SOUPS 2019)},
    year = {2019},
    isbn = {978-1-939133-05-2},
    address = {Santa Clara, CA},
    pages = {97--115},
    url = {https://www.usenix.org/conference/soups2019/presentation/das},
    publisher = {USENIX Association},
    month = aug
}

@inproceedings {conf/soups/DiMartinoRWQLA19,
    author = {Di Martino, Mariano and Pieter Robyns and Winnie Weyts and Peter Quax and Wim Lamotte and Ken Andries},
    title = {Personal Information Leakage by Abusing the {GDPR} {\textquoteright}Right of Access{\textquoteright}},
    booktitle = {Fifteenth Symposium on Usable Privacy and Security (SOUPS 2019)},
    year = {2019},
    isbn = {978-1-939133-05-2},
    address = {Santa Clara, CA},
    pages = {371--385},
    url = {https://www.usenix.org/conference/soups2019/presentation/dimartino},
    publisher = {USENIX Association},
    month = aug
}

@inproceedings {conf/soups/DruryM19,
    author = {Vincent Drury and Ulrike Meyer},
    title = {Certified Phishing: Taking a Look at Public Key Certificates of Phishing Websites},
    booktitle = {Fifteenth Symposium on Usable Privacy and Security (SOUPS 2019)},
    year = {2019},
    isbn = {978-1-939133-05-2},
    address = {Santa Clara, CA},
    pages = {211--223},
    url = {https://www.usenix.org/conference/soups2019/presentation/drury},
    publisher = {USENIX Association},
    month = aug
}

@inproceedings {conf/soups/FaklarisDH19,
    author = {Cori Faklaris and Laura A. Dabbish and Jason I. Hong},
    title = {A {Self-Report} Measure of {End-User} Security Attitudes ({{{{{SA-6}}}}})},
    booktitle = {Fifteenth Symposium on Usable Privacy and Security (SOUPS 2019)},
    year = {2019},
    isbn = {978-1-939133-05-2},
    address = {Santa Clara, CA},
    pages = {61--77},
    url = {https://www.usenix.org/conference/soups2019/presentation/faklaris},
    publisher = {USENIX Association},
    month = aug
}

@inproceedings {conf/soups/FrikNBLSE19,
    author = {Alisa Frik and Leysan Nurgalieva and Julia Bernd and Joyce Lee and Florian Schaub and Serge Egelman},
    title = {Privacy and Security Threat Models and Mitigation Strategies of Older Adults},
    booktitle = {Fifteenth Symposium on Usable Privacy and Security (SOUPS 2019)},
    year = {2019},
    isbn = {978-1-939133-05-2},
    address = {Santa Clara, CA},
    pages = {21--40},
    url = {https://www.usenix.org/conference/soups2019/presentation/frik},
    publisher = {USENIX Association},
    month = aug
}

@inproceedings {conf/soups/FultonGMARM19,
    author = {Kelsey R. Fulton and Rebecca Gelles and Alexandra McKay and Yasmin Abdi and Richard Roberts and Michelle L. Mazurek},
    title = {The Effect of Entertainment Media on Mental Models of Computer Security},
    booktitle = {Fifteenth Symposium on Usable Privacy and Security (SOUPS 2019)},
    year = {2019},
    isbn = {978-1-939133-05-2},
    address = {Santa Clara, CA},
    pages = {79--95},
    url = {https://www.usenix.org/conference/soups2019/presentation/fulton},
    publisher = {USENIX Association},
    month = aug
}

@inproceedings {conf/soups/HabibZJSSACSS19,
    author = {Hana Habib and Yixin Zou and Aditi Jannu and Neha Sridhar and Chelse Swoopes and Alessandro Acquisti and Lorrie Faith Cranor and Norman Sadeh and Florian Schaub},
    title = {An Empirical Analysis of Data Deletion and {Opt-Out} Choices on 150 Websites},
    booktitle = {Fifteenth Symposium on Usable Privacy and Security (SOUPS 2019)},
    year = {2019},
    isbn = {978-1-939133-05-2},
    address = {Santa Clara, CA},
    pages = {387--406},
    url = {https://www.usenix.org/conference/soups2019/presentation/habib},
    publisher = {USENIX Association},
    month = aug
}

@inproceedings {conf/soups/HayesKPW19,
    author = {Jordan Hayes and Smirity Kaushik and Charlotte Emily Price and Yang Wang},
    title = {Cooperative Privacy and Security: Learning from People with Visual Impairments and Their Allies},
    booktitle = {Fifteenth Symposium on Usable Privacy and Security (SOUPS 2019)},
    year = {2019},
    isbn = {978-1-939133-05-2},
    address = {Santa Clara, CA},
    pages = {1--20},
    url = {https://www.usenix.org/conference/soups2019/presentation/hayes},
    publisher = {USENIX Association},
    month = aug
}

@inproceedings {conf/soups/KumRIRLS19,
    author = {Hye-Chung Kum and Eric D. Ragan and Gurudev Ilangovan and Mahin Ramezani and Qinbo Li and Cason Schmit},
    title = {Enhancing Privacy through an Interactive On-demand Incremental Information Disclosure Interface: Applying {Privacy-by-Design} to Record Linkage},
    booktitle = {Fifteenth Symposium on Usable Privacy and Security (SOUPS 2019)},
    year = {2019},
    isbn = {978-1-939133-05-2},
    address = {Santa Clara, CA},
    pages = {175--189},
    url = {https://www.usenix.org/conference/soups2019/presentation/kum},
    publisher = {USENIX Association},
    month = aug
}

@inproceedings {conf/soups/LiRMMC19,
    author = {Frank Li and Lisa Rogers and Arunesh Mathur and Nathan Malkin and Marshini Chetty},
    title = {Keepers of the Machines: Examining How System Administrators Manage Software Updates For Multiple Machines},
    booktitle = {Fifteenth Symposium on Usable Privacy and Security (SOUPS 2019)},
    year = {2019},
    isbn = {978-1-939133-05-2},
    address = {Santa Clara, CA},
    pages = {273--288},
    url = {https://www.usenix.org/conference/soups2019/presentation/li},
    publisher = {USENIX Association},
    month = aug
}

@inproceedings {conf/soups/MeckeBKPA19,
    author = {Lukas Mecke and Daniel Buschek and Mathias Kiermeier and Sarah Prange and Florian Alt},
    title = {Exploring Intentional Behaviour Modifications for Password Typing on Mobile Touchscreen Devices},
    booktitle = {Fifteenth Symposium on Usable Privacy and Security (SOUPS 2019)},
    year = {2019},
    isbn = {978-1-939133-05-2},
    address = {Santa Clara, CA},
    pages = {303--317},
    url = {https://www.usenix.org/conference/soups2019/presentation/mecke-behaviour},
    publisher = {USENIX Association},
    month = aug
}

@inproceedings {conf/soups/MeckeRBPA19,
    author = {Lukas Mecke and Sarah Delgado Rodriguez and Daniel Buschek and Sarah Prange and Florian Alt},
    title = {Communicating Device Confidence Level and Upcoming {Re-Authentications} in Continuous Authentication Systems on Mobile Devices},
    booktitle = {Fifteenth Symposium on Usable Privacy and Security (SOUPS 2019)},
    year = {2019},
    isbn = {978-1-939133-05-2},
    address = {Santa Clara, CA},
    pages = {289--301},
    url = {https://www.usenix.org/conference/soups2019/presentation/mecke-confidence},
    publisher = {USENIX Association},
    month = aug
}

@inproceedings {conf/soups/MhaidliZS19,
    author = {Abraham H. Mhaidli and Yixin Zou and Florian Schaub},
    title = {"We Can{\textquoteright}t Live Without {Them!}" App Developers{\textquoteright} Adoption of Ad Networks and Their Considerations of Consumer Risks},
    booktitle = {Fifteenth Symposium on Usable Privacy and Security (SOUPS 2019)},
    year = {2019},
    isbn = {978-1-939133-05-2},
    address = {Santa Clara, CA},
    pages = {225--244},
    url = {https://www.usenix.org/conference/soups2019/presentation/mhaidli},
    publisher = {USENIX Association},
    month = aug
}

@inproceedings {conf/soups/PatnaikHR19,
    author = {Nikhil Patnaik and Joseph Hallett and Awais Rashid},
    title = {Usability Smells: An Analysis of {Developers{\textquoteright}} Struggle With Crypto Libraries},
    booktitle = {Fifteenth Symposium on Usable Privacy and Security (SOUPS 2019)},
    year = {2019},
    isbn = {978-1-939133-05-2},
    address = {Santa Clara, CA},
    pages = {245--257},
    url = {https://www.usenix.org/conference/soups2019/presentation/patnaik},
    publisher = {USENIX Association},
    month = aug
}

@inproceedings {conf/soups/PearmanZBCC19,
    author = {Sarah Pearman and Shikun Aerin Zhang and Lujo Bauer and Nicolas Christin and Lorrie Faith Cranor},
    title = {Why people (don{\textquoteright}t) use password managers effectively},
    booktitle = {Fifteenth Symposium on Usable Privacy and Security (SOUPS 2019)},
    year = {2019},
    isbn = {978-1-939133-05-2},
    address = {Santa Clara, CA},
    pages = {319--338},
    url = {https://www.usenix.org/conference/soups2019/presentation/pearman},
    publisher = {USENIX Association},
    month = aug
}

@inproceedings {conf/soups/QinLJFAGRV19,
    author = {Lucy Qin and Andrei Lapets and Frederick Jansen and Peter Flockhart and Kinan Dak Albab and Ira Globus-Harris and Shannon Roberts and Mayank Varia},
    title = {From Usability to Secure Computing and Back Again},
    booktitle = {Fifteenth Symposium on Usable Privacy and Security (SOUPS 2019)},
    year = {2019},
    isbn = {978-1-939133-05-2},
    address = {Santa Clara, CA},
    pages = {191--210},
    url = {https://www.usenix.org/conference/soups2019/presentation/qin},
    publisher = {USENIX Association},
    month = aug
}

@inproceedings {conf/soups/ReeseSDACS19,
    author = {Ken Reese and Trevor Smith and Jonathan Dutson and Jonathan Armknecht and Jacob Cameron and Kent Seamons},
    title = {A Usability Study of Five {Two-Factor} Authentication Methods},
    booktitle = {Fifteenth Symposium on Usable Privacy and Security (SOUPS 2019)},
    year = {2019},
    isbn = {978-1-939133-05-2},
    address = {Santa Clara, CA},
    pages = {357--370},
    url = {https://www.usenix.org/conference/soups2019/presentation/reese},
    publisher = {USENIX Association},
    month = aug
}

@inproceedings {conf/soups/SimoiuBGG19,
    author = {Camelia Simoiu and Joseph Bonneau and Christopher Gates and Sharad Goel},
    title = {"I was told to buy a software or lose my computer. I ignored it": A study of ransomware},
    booktitle = {Fifteenth Symposium on Usable Privacy and Security (SOUPS 2019)},
    year = {2019},
    isbn = {978-1-939133-05-2},
    address = {Santa Clara, CA},
    pages = {155--174},
    url = {https://www.usenix.org/conference/soups2019/presentation/simoiu},
    publisher = {USENIX Association},
    month = aug
}

@inproceedings {conf/soups/TabassumKL19,
    author = {Madiha Tabassum and Tomasz Kosinski and Heather Richter Lipford},
    title = {"I don{\textquoteright}t own the data": End User Perceptions of Smart Home Device Data Practices and Risks},
    booktitle = {Fifteenth Symposium on Usable Privacy and Security (SOUPS 2019)},
    year = {2019},
    isbn = {978-1-939133-05-2},
    address = {Santa Clara, CA},
    pages = {435--450},
    url = {https://www.usenix.org/conference/soups2019/presentation/tabassum},
    publisher = {USENIX Association},
    month = aug
}

@inproceedings {conf/soups/VanceEJKA19,
    author = {Anthony Vance and David Eargle and Jeffrey L. Jenkins and C. Brock Kirwan and Bonnie Brinton Anderson},
    title = {The Fog of Warnings: How Non-essential Notifications Blur with Security Warnings},
    booktitle = {Fifteenth Symposium on Usable Privacy and Security (SOUPS 2019)},
    year = {2019},
    isbn = {978-1-939133-05-2},
    address = {Santa Clara, CA},
    pages = {407--420},
    url = {https://www.usenix.org/conference/soups2019/presentation/vance},
    publisher = {USENIX Association},
    month = aug
}

@inproceedings {conf/soups/VoronkovML19,
    author = {Artem Voronkov and Leonardo A. Martucci and Stefan Lindskog},
    title = {System Administrators Prefer Command Line Interfaces, Don{\textquoteright}t They? An Exploratory Study of Firewall Interfaces},
    booktitle = {Fifteenth Symposium on Usable Privacy and Security (SOUPS 2019)},
    year = {2019},
    isbn = {978-1-939133-05-2},
    address = {Santa Clara, CA},
    pages = {259--271},
    url = {https://www.usenix.org/conference/soups2019/presentation/voronkov},
    publisher = {USENIX Association},
    month = aug
}

@inproceedings {conf/soups/WuGHTVZS19,
    author = {Justin Wu and Cyrus Gattrell and Devon Howard and Jake Tyler and Elham Vaziripour and Daniel Zappala and Kent Seamons},
    title = {"Something isn{\textquoteright}t secure, but I{\textquoteright}m not sure how that translates into a problem": Promoting autonomy by designing for understanding in Signal},
    booktitle = {Fifteenth Symposium on Usable Privacy and Security (SOUPS 2019)},
    year = {2019},
    isbn = {978-1-939133-05-2},
    address = {Santa Clara, CA},
    pages = {137--153},
    url = {https://www.usenix.org/conference/soups2019/presentation/wu},
    publisher = {USENIX Association},
    month = aug
}

@article{murthy2021individually,
  title={Individually vulnerable, collectively safe: The security and privacy practices of households with older adults},
  author={Murthy, Savanthi and Bhat, Karthik S and Das, Sauvik and Kumar, Neha},
  journal={Proceedings of the ACM on Human-Computer Interaction},
  volume={5},
  number={CSCW1},
  pages={1--24},
  year={2021},
  publisher={ACM New York, NY, USA}
}

@conference {teradareplication,
title = {Replication: Application of Security Attitudes Scale to Japanese Workers},
author = {Terada, Takeaki and Furukawa, Kazuyoshi},
year = {2021},
publisher = {USENIX Association},
month = aug
}

@inproceedings{berridge2022control,
  title={Control Matters in Elder Care Technology: Evidence and Direction for Designing It In},
  author={Berridge, Clara and Zhou, Yuanjin and Lazar, Amanda and Porwal, Anupreet and Mattek, Nora and Gothard, Sarah and Kaye, Jeffrey},
  booktitle={Designing Interactive Systems Conference},
  pages={1831--1848},
  year={2022}
}

@article{ray2020warn,
  title={“Warn Them” or “Just Block Them”?: Investigating Privacy Concerns Among Older and Working Age Adults},
  author={Ray, Hirak and Wolf, Flynn and Kuber, Ravi and Aviv, Adam J},
  journal={UMBC Student Collection},
  year={2020},
  publisher={Sciendo}
}

@inproceedings {conf/soups/TiefenauHKZ20,
author = {Christian Tiefenau and Maximilian H{\"a}ring and Katharina Krombholz and Emanuel von Zezschwitz},
title = {Security, Availability, and Multiple Information Sources: Exploring Update Behavior of System Administrators},
booktitle = {Sixteenth Symposium on Usable Privacy and Security (SOUPS 2020)},
year = {2020},
isbn = {978-1-939133-16-8},
pages = {239--258},
url = {https://www.usenix.org/conference/soups2020/presentation/tiefenau},
publisher = {USENIX Association},
month = aug
}

@inproceedings{martius2020does,
  title={What does this update do to my systems?--An analysis of the importance of update-related information to system administrators},
  author={Martius, Florin and Tiefenau, Christian},
  booktitle={Workshop on Security Information Workers, WSIW},
  volume={20},
  pages={1--12},
  year={2020}
}

@inproceedings {conf/soups/MayerMMA22,
author = {Peter Mayer and Collins W. Munyendo and Michelle L. Mazurek and Adam J. Aviv},
title = {Why Users (Don{\textquoteright}t) Use Password Managers at a Large Educational Institution},
booktitle = {31st USENIX Security Symposium (USENIX Security 22)},
year = {2022},
isbn = {978-1-939133-31-1},
address = {Boston, MA},
pages = {1849--1866},
url = {https://www.usenix.org/conference/usenixsecurity22/presentation/mayer},
publisher = {USENIX Association},
month = aug
}

@misc{sereda2022supporting,
  title={Supporting End Users in Securing IoT-Enabled Smart Home Devices.},
  author={Sereda, Bohdana},
  year={2022},
  school={Carleton University}
}

@article{kruzikova2022usable,
  title={Usable and secure? User perception of four authentication methods for mobile banking},
  author={Kruzikova, Agata and Knapova, Lenka and Smahel, David and Dedkova, Lenka and Matyas, Vashek},
  journal={Computers \& Security},
  volume={115},
  pages={102603},
  year={2022},
  publisher={Elsevier}
}

@online{soups-proceedings,
    title = {SOUPS 2019 Technical Sessions},
    year = {2019},
    url = {https://www.usenix.org/conference/soups2019/technical-sessions},
    urldate = {2024-02-12}
}

@online{dblp,
    title = {dblp computer science bibliography},
    year = {2024},
    url = {https://dblp.org/},
    urldate = {2024-02-12}
}

@article{conf/asist/DingZCSWZ14,
author = {Ding, Ying and Zhang, Guo and Chambers, Tamy and Song, Min and Wang, Xiaolong and Zhai, Chengxiang},
title = {Content-based citation analysis: The next generation of citation analysis},
journal = {Journal of the Association for Information Science and Technology},
volume = {65},
number = {9},
pages = {1820-1833},
doi = {https://doi.org/10.1002/asi.23256},
url = {https://asistdl.onlinelibrary.wiley.com/doi/abs/10.1002/asi.23256},
eprint = {https://asistdl.onlinelibrary.wiley.com/doi/pdf/10.1002/asi.23256},
year = {2014}
}

@article{journal/NLE/JhaJQR17, 
title={NLP-driven citation analysis for scientometrics}, volume={23}, 
DOI={10.1017/S1351324915000443}, 
number={1}, 
journal={Natural Language Engineering}, 
author={Jha, Rahul and Jbara, Amjad-Abu and Qazvinian, Vahed and Radev, Dragomir R.}, 
year={2017}, 
pages={93–130}
}

@inproceedings{conf/sigir/Teufel17,
  author       = {Simone Teufel},
  editor       = {Philipp Mayr and
                  Muthu Kumar Chandrasekaran and
                  Kokil Jaidka},
  title        = {Do "Future Work" sections have a purpose? Citation links and entailment
                  for global scientometric questions},
  booktitle    = {Proceedings of the 2nd Joint Workshop on Bibliometric-enhanced Information
                  Retrieval and Natural Language Processing for Digital Libraries {(BIRNDL}
                  2017) co-located with the 40th International {ACM} {SIGIR} Conference
                  on Research and Development in Information Retrieval {(SIGIR} 2017),
                  Tokyo, Japan, August 11, 2017},
  series       = {{CEUR} Workshop Proceedings},
  volume       = {1888},
  pages        = {7--13},
  publisher    = {CEUR-WS.org},
  year         = {2017},
  url          = {https://ceur-ws.org/Vol-1888/paper1.pdf},
  bibsource    = {dblp computer science bibliography, https://dblp.org}
}

@inproceedings{conf/issi/LiY19,
  author       = {Kai Li and
                  Erjia Yan},
  editor       = {Giuseppe Catalano and
                  Cinzia Daraio and
                  Martina Gregori and
                  Henk F. Moed and
                  Giancarlo Ruocco},
  title        = {Using a keyword extraction pipeline to understand concepts in future
                  work sections of research papers},
  booktitle    = {Proceedings of the 17th International Conference on Scientometrics
                  and Informetrics, {ISSI} 2019, Rome, Italy, September 2-5, 2019},
  pages        = {87--98},
  publisher    = {{ISSI} Society},
  year         = {2019},
  bibsource    = {dblp computer science bibliography, https://dblp.org}
}

@inproceedings{conf/jcdl/HaoLQWZ20,
author = {Hao, Wenke and Li, Zhicheng and Qian, Yuchen and Wang, Yuzhuo and Zhang, Chengzhi},
title = {The ACL FWS-RC: A Dataset for Recognition and Classification of Sentence about Future Works},
year = {2020},
isbn = {9781450375856},
publisher = {Association for Computing Machinery},
address = {New York, NY, USA},
url = {https://doi.org/10.1145/3383583.3398526},
doi = {10.1145/3383583.3398526},
booktitle = {Proceedings of the ACM/IEEE Joint Conference on Digital Libraries in 2020},
pages = {261–269},
numpages = {9},
location = {Virtual Event, China},
series = {JCDL '20}
}

@misc{arxiv/MarshalovaBB23,
      title={Automatic Aspect Extraction from Scientific Texts}, 
      author={Anna Marshalova and Elena Bruches and Tatiana Batura},
      year={2023},
      eprint={2310.04074},
      archivePrefix={arXiv},
      primaryClass={cs.CL}
}

@article{journal/informetrics/ZhangXHLQW23,
title = {Automatic recognition and classification of future work sentences from academic articles in a specific domain},
journal = {Journal of Informetrics},
volume = {17},
number = {1},
pages = {101373},
year = {2023},
issn = {1751-1577},
doi = {https://doi.org/10.1016/j.joi.2022.101373},
url = {https://www.sciencedirect.com/science/article/pii/S1751157722001262},
author = {Chengzhi Zhang and Yi Xiang and Wenke Hao and Zhicheng Li and Yuchen Qian and Yuzhuo Wang}
}

@article{journals/corr/HuW15a,
  author       = {Yue Hu and
                  Xiaojun Wan},
  title        = {Mining and Analyzing the Future Works in Scientific Articles},
  journal      = {CoRR},
  volume       = {abs/1507.02140},
  year         = {2015},
  url          = {http://arxiv.org/abs/1507.02140},
  eprinttype    = {arXiv},
  eprint       = {1507.02140},
  bibsource    = {dblp computer science bibliography, https://dblp.org}
}

@article{conf/asist/QianLHWZ21,
author = {Qian, Yuchen and Li, Zhicheng and Hao, Wenke and Wang, Yuzhuo and Zhang, Chengzhi},
title = {Using Future Work Sentences to Explore Research Trends of Different Tasks in a Special Domain},
journal = {Proceedings of the Association for Information Science and Technology},
volume = {58},
number = {1},
pages = {532-536},
doi = {https://doi.org/10.1002/pra2.492},
url = {https://asistdl.onlinelibrary.wiley.com/doi/abs/10.1002/pra2.492},
eprint = {https://asistdl.onlinelibrary.wiley.com/doi/pdf/10.1002/pra2.492},
year = {2021}
}

@article{journal/DADK/RuoxuanLY21,
author = {Song Ruoxuan,Qian Li,Du Yu},
title = {Identifying Academic Creative Concept Topics Based on Future Work of Scientific Papers},
publisher = {Data Analysis and Knowledge Discovery},
year = {2021},
journal = {Data Analysis and Knowledge Discovery},
volume = {5},
number = {5},
eid = {10},
numpages = {10},
pages = {10},
url = {https://manu44.magtech.com.cn/Jwk_infotech_wk3/EN/abstract/article_5059.shtml},
doi = {10.11925/infotech.2096-3467.2020.1275}
}

@article{journal/scientometrics/ChaoCZL23,
author = {Chao, Wenhan and Chen, Mengyuan and Zhou, Xian and Luo, Zhunchen},
title = {A joint framework for identifying the type and arguments of scientific contribution},
journal = {Scientometrics},
volume = {128},
doi = {https://doi.org/10.1007/s11192-023-04694-6},
year = {2023}
}

@article{conf/asist/ZhuWS19,
author = {Zhu, Zihe and Wang, Dongbo and Shen, Si},
title = {Recognizing sentences concerning future research from the full text of JASIST},
journal = {Proceedings of the Association for Information Science and Technology},
volume = {56},
number = {1},
pages = {858-859},
doi = {https://doi.org/10.1002/pra2.206},
url = {https://asistdl.onlinelibrary.wiley.com/doi/abs/10.1002/pra2.206},
eprint = {https://asistdl.onlinelibrary.wiley.com/doi/pdf/10.1002/pra2.206},
year = {2019}
}

@article{journal/DTA/DehdariradGJ20,
author = {Dehdarirad, Hossein and Ghazimirsaeid, Javad and Jalalimanesh, Ammar},
title = {A joint framework for identifying the type and arguments of scientific contribution},
journal = {Data Technologies and Applications},
volume = {54},
doi = {https://doi.org/10.1108/DTA-08-2019-0135},
year = {2020},
publisher = {Emerald Publishing Limited}
}

@article{journal/PAJAIS/BarataC23,
author = {João Barata and Paolo Rupino da Cunha},
title = {Getting Around to It: How Design Science Researchers Set Future Work Agendas},
journal = {PAJAIS Preprints},
year = {2023},
}

@article{journal/lancet/Malterud01,
    title = {Qualitative research: standards, challenges, and guidelines},
    journal = {The Lancet},
    volume = {358},
    number = {9280},
    pages = {483-488},
    year = {2001},
    doi = {https://doi.org/10.1016/S0140-6736(01)05627-6},
    author = {Kirsti Malterud},
}

@INPROCEEDINGS{conf/ssp/LadisaPMB23,
author = {Piergiorgio Ladisa and Henrik Plate and Matias Martinez and Olivier Barais},
booktitle = {2023 IEEE Symposium on Security and Privacy (SP)},
title = {SoK: Taxonomy of Attacks on Open-Source Software Supply Chains},
year = {2023},
pages = {1509--1526},
doi = {10.1109/SP46215.2023.10179304},
publisher = {IEEE},
month = {05}
}

@article{keshav2007read,
  title={How to read a paper},
  author={Keshav, Srinivasan},
  journal={ACM SIGCOMM Computer Communication Review},
  volume={37},
  number={3},
  pages={83--84},
  year={2007},
  publisher={ACM New York, NY, USA}
}

@inproceedings{10.1145/3411764.3445122,
author = {Abdi, Noura and Zhan, Xiao and Ramokapane, Kopo M. and Such, Jose},
title = {Privacy Norms for Smart Home Personal Assistants},
year = {2021},
isbn = {9781450380966},
publisher = {Association for Computing Machinery},
address = {New York, NY, USA},
url = {https://doi.org/10.1145/3411764.3445122},
doi = {10.1145/3411764.3445122},
booktitle = {Proceedings of the 2021 CHI Conference on Human Factors in Computing Systems},
articleno = {558},
numpages = {14},
location = {, Yokohama, Japan, },
series = {CHI '21}
}

@inproceedings {conf/usenix/AbrokwaDAM21,
author = {Desiree Abrokwa and Shruti Das and Omer Akgul and Michelle L. Mazurek},
title = {Comparing Security and Privacy Attitudes Among {U.S}. Users of Different Smartphone and {Smart-Speaker} Platforms},
booktitle = {Seventeenth Symposium on Usable Privacy and Security (SOUPS 2021)},
year = {2021},
isbn = {978-1-939133-25-0},
pages = {139--158},
url = {https://www.usenix.org/conference/soups2021/presentation/abrokwa},
publisher = {USENIX Association},
month = aug
}

@INPROCEEDINGS{PaperID541,
  author={Homayoun, Sajad and Hageman, Kaspar and Afzal-Houshmand, Sam and Jensen, Christian D. and Pedersen, Jens M.},
  booktitle={2022 International Conference on Information Networking (ICOIN)}, 
  title={Detecting Ambiguous Phishing Certificates using Machine Learning}, 
  year={2022},
  volume={},
  number={},
  pages={1-6},
  doi={10.1109/ICOIN53446.2022.9687264}}

@article{PaperID544,
  title={Exploring a robust machine learning classifier for detecting phishing domains using SSL certificates},
  author={Akanchha, Akanchha},
  year={2020}
}

@article{PaperID1200,
  title={Linked by age: a study on social media privacy concerns among younger and older adults},
  author={Goyeneche, David and Singaraju, Stephen and Arango, Luis},
  journal={Industrial Management \& Data Systems},
  volume={124},
  number={2},
  pages={640--665},
  year={2024},
  publisher={Emerald Publishing Limited}
}

@inproceedings{PaperID355,
author = {Ortloff, Anna-Marie and Vossen, Maike and Tiefenau, Christian},
title = {Replicating a Study of Ransomware in Germany},
year = {2021},
isbn = {9781450384230},
publisher = {Association for Computing Machinery},
address = {New York, NY, USA},
url = {https://doi.org/10.1145/3481357.3481508},
doi = {10.1145/3481357.3481508},
booktitle = {Proceedings of the 2021 European Symposium on Usable Security},
pages = {151–164},
numpages = {14},
location = {Karlsruhe, Germany},
series = {EuroUSEC '21}
}

@inproceedings {PaperID711,
author = {Peter Mayer and Collins W. Munyendo and Michelle L. Mazurek and Adam J. Aviv},
title = {Why Users (Don{\textquoteright}t) Use Password Managers at a Large Educational Institution},
booktitle = {31st USENIX Security Symposium (USENIX Security 22)},
year = {2022},
isbn = {978-1-939133-31-1},
address = {Boston, MA},
pages = {1849--1866},
url = {https://www.usenix.org/conference/usenixsecurity22/presentation/mayer},
publisher = {USENIX Association},
month = aug
}

@inproceedings {PaperID706,
author = {Hirak Ray and Flynn Wolf and Ravi Kuber and Adam J. Aviv},
title = {Why Older Adults (Don{\textquoteright}t) Use Password Managers},
booktitle = {30th USENIX Security Symposium (USENIX Security 21)},
year = {2021},
isbn = {978-1-939133-24-3},
pages = {73--90},
url = {https://www.usenix.org/conference/usenixsecurity21/presentation/ray},
publisher = {USENIX Association},
month = aug
}

@article{PaperID631,
  title={Revisiting identification issues in GDPR ‘Right Of Access’ policies: a technical and longitudinal analysis},
  author={Di Martino, Mariano and Meers, Isaac and Quax, Peter and Andries, Ken and Lamotte, Wim},
  journal={Proceedings on Privacy Enhancing Technologies},
  volume={2022},
  number={2},
  pages={95--113},
  year={2022}
}

@inproceedings {baigreplication,
author = {Khadija Baig and Elisa Kazan and Kalpana Hundlani and Sana Maqsood and Sonia Chiasson},
title = {Replication: Effects of Media on the Mental Models of Technical Users},
booktitle = {Seventeenth Symposium on Usable Privacy and Security (SOUPS 2021)},
year = {2021},
isbn = {978-1-939133-25-0},
pages = {119--138},
url = {https://www.usenix.org/conference/soups2021/presentation/baig},
publisher = {USENIX Association},
month = aug
}

@inproceedings{haney2020smart,
  title={Smart home security and privacy mitigations: Consumer perceptions, practices, and challenges},
  author={Haney, Julie M and Furman, Susanne M and Acar, Yasemin},
  booktitle={HCI for Cybersecurity, Privacy and Trust: Second International Conference, HCI-CPT 2020, Held as Part of the 22nd HCI International Conference, HCII 2020, Copenhagen, Denmark, July 19--24, 2020, Proceedings 22},
  pages={393--411},
  year={2020},
  organization={Springer}
}

@inproceedings {PaperID378,
author = {Maximilian Golla and Grant Ho and Marika Lohmus and Monica Pulluri and Elissa M. Redmiles},
title = {Driving {2FA} Adoption at Scale: Optimizing {Two-Factor} Authentication Notification Design Patterns},
booktitle = {30th USENIX Security Symposium (USENIX Security 21)},
year = {2021},
isbn = {978-1-939133-24-3},
pages = {109--126},
url = {https://www.usenix.org/conference/usenixsecurity21/presentation/golla},
publisher = {USENIX Association},
month = aug
}

@article{PaperID181,
  title={Third-party web tracking under the general data protection regulation},
  author={Utz, Christine},
  year={2023}
}

@inproceedings{PaperID1202,
  title={To disclose or not to disclose: examining the privacy decision-making processes of older vs. younger adults},
  author={Ghaiumy Anaraky, Reza and Byrne, Kaileigh Angela and Wisniewski, Pamela J and Page, Xinru and Knijnenburg, Bart},
  booktitle={Proceedings of the 2021 CHI Conference on Human Factors in Computing Systems},
  pages={1--14},
  year={2021}
}

@inproceedings{PaperID538,
author = {AlSabah, Mashael and Nabeel, Mohamed and Boshmaf, Yazan and Choo, Euijin},
title = {Content-Agnostic Detection of Phishing Domains using Certificate Transparency and Passive DNS},
year = {2022},
isbn = {9781450397049},
publisher = {Association for Computing Machinery},
address = {New York, NY, USA},
url = {https://doi.org/10.1145/3545948.3545958},
doi = {10.1145/3545948.3545958},
booktitle = {Proceedings of the 25th International Symposium on Research in Attacks, Intrusions and Defenses},
pages = {446–459},
numpages = {14},
location = {Limassol, Cyprus},
series = {RAID '22}
}

@online{soups-cfp,
    title = {SOUPS 2024 Call for Papers | USENIX},
    url = {https://www.usenix.org/conference/soups2024/call-for-papers},
    urldate = {2024-02-15},
}

@string{acm = {ACM}}

@string{ieee = {IEEE}}

@string{usenix = {USENIX}}

@string{springer = {Springer}}

@string{elsevier = {Elsevier B.V.}}

@periodical{cscw,
 journal = {Computer Supported Cooperative Work (CSCW)},
 publisher = {Springer Netherlands}
}

@periodical{sp,
 journal = {IEEE Security \& Privacy },
 publisher = {IEEE}
}

\end{document}